\documentclass[prd,aps,nofootinbib,floatfix,11pt]{revtex4}
\usepackage{amsmath,graphicx,epsfig,amssymb,dsfont,mathtools,}
\usepackage[usenames]{color}
\usepackage{ulem} 
\usepackage{bigstrut}
\usepackage{slashed}
\usepackage{multirow}
\usepackage{subfigure}

\allowdisplaybreaks

\begin{document}\title{Towards a Heavy Diquark Effective Theory for  Weak Decays\\ of Doubly Heavy Baryons}
\author{Yu-Ji Shi~$^{1,2}$~\footnote{Email:shiyuji92@126.com},
  Wei Wang~$^{2}$~\footnote{Email:wei.wang@sjtu.edu.cn},
  Zhen-Xing Zhao~$^{2,3}$~\footnote{Email:star\_0027@sjtu.edu.cn},
  Ulf-G. Mei{\ss}ner~$^{1,4,5}$~\footnote{Email:meissner@hiskp.uni-bonn.de}}

\affiliation{$^{1}$ Helmholtz-Institut f\"ur Strahlen- und Kernphysik and Bethe Center \\ for Theoretical Physics,
  Universit\"at Bonn, 53115 Bonn, Germany\\
$^{2}$ INPAC, SKLPPC, MOE KLPPC, School of Physics and Astronomy, Shanghai Jiao Tong University, Shanghai 200240, China\\
$^{3}$ School of Physical Science and Technology, Inner Mongolia University, Hohhot 010021, China\\
  $^4$ Institute for Advanced Simulation, Institut f{\"u}r Kernphysik and J\"ulich Center for Hadron Physics,
  Forschungszentrum J{\"u}lich, D-52425 J{\"u}lich, Germany\\
$^5$ Tbilisi State University, 0186 Tbilisi, Georgia}

\begin{abstract}
We construct a leading-order effective field theory for both scalar and axial-vector heavy diquarks, and consider
its power expansion in the heavy diquark limit. By assuming the transition from QCD to diquark effective
theory, we derive  the most general form for the effective diquark transition currents based on the heavy
diquark symmetry. The short-distance coefficients between QCD and heavy diquark effective field theory
are also obtained by a tree level matching. With the effective currents in the heavy diquark limit,
we perform a reduction of the form factors for semi-leptonic decays of doubly heavy baryons,  and find that
only one nonperturbative function is remaining. It is shown that this soft function can be related to
the Isgur-Wise function in heavy meson transitions.  As a phenomenological application, we take  a
single pole structure for the reduced form factor, and  use it to calculate  the semi-leptonic decay
widths of doubly heavy baryons. The obtained results are consistent with others given in the literature,
and can be tested in the future.
\end{abstract}
\maketitle

\section{Introduction}
In the past, the conventionl quark model has successfully   explained   structures of
numerous hadronic states observed in a large number
of experiments. However, not all predicted particles by the quark model have been
experimentally established. In particular, doubly heavy baryons, that is baryonic states made of two
heavy quarks, are  of this type. After pursuing the $\Xi_{cc}$ for many years,
the LHCb collaboration finally announced in 2017  the
observation of $\Xi_{cc}^{++}$, a lowest-lying doubly-charmed baryon whose mass is give as~\cite{Aaij:2017ueg}
\begin{equation}
m_{\Xi_{cc}^{++}}=(3621.40\pm0.72\pm0.27\pm0.14)\ {\rm MeV}.\label{eq:LHCb_measurement}
\end{equation}
This inspiring observation follows  an earlier prediction Ref.~\cite{Yu:2017zst},
where  the $\Xi_{cc}^{++}$ is expected to be reconstructed from the decay channel
$\Xi_{cc}^{++}\to\Lambda_{c}^{+}K^{-}\pi^{+}\pi^{+}$. One year later,  LHCb has also successfully
measured the $\Xi_{cc}^{++}$'s lifetime~\cite{Aaij:2018wzf}, and   reconstructed this resonance
from the $\Xi_{c}^{+}\pi^{+}$ final state~\cite{Aaij:2018gfl}. Thus, the existence of the $\Xi_{cc}^{++}$ is
unambiguously established. We believe that  through   continuous experimental
efforts~\cite{Traill:2017zbs,Cerri:2018ypt,Aaij:2019jfq}, other heavier doubly heavy baryons could
be discovered in the future. In addition, there have been numerous theoretical studies aiming to
understand the dynamical and spectroscopical
properties of the doubly-heavy baryon states, see e.g. Refs.~\cite{Wang:2017mqp,Wang:2017azm,
Gutsche:2017hux,Li:2017pxa,Guo:2017vcf,Xiao:2017udy,Sharma:2017txj,Ma:2017nik,Hu:2017dzi,Shi:2017dto,Yao:2018zze,Yao:2018ifh,
Ozdem:2018uue,Ali:2018ifm,Zhao:2018mrg,Wang:2018lhz,
Liu:2018euh,Xing:2018lre,Dhir:2018twm,Berezhnoy:2018bde,Jiang:2018oak,Zhang:2018llc,Li:2018bkh,Gutsche:2018msz,Shi:2019hbf,Shi:2019fph,Hu:2019bqj,Brodsky:2011zs,Yan:2018zdt,Hu:2020mxk}. However, a comprehensive description of
the decay mechanism of doubly heavy baryons is not established yet. 

Generally, an ideal platform for studying hadrons is through semi-leptonic weak decays. The main advantage of a
semi-leptonic process is  its naturalness  in separating the QCD relevant and the QCD irrelevant dynamics in
the weak decays. All the QCD dynamics is encapsulated in the hadron transition matrix element, which is
independent from the leptonic part and  can be parametrized by several
form factors. However, as a three-body system, a doubly heavy baryon posseses a
much more complicated dynamics than a  heavy meson.

A straightforward way to consider this problem is to reduce a doubly heavy baryon into  a two-body system,
where two of the three quarks are treated as a point-like diquarks.  Generally, each two quarks in a
baryon form a color antitriplet so that they might be  bound  by
an attractive potential. However, for a doubly heavy baryon, it is more reasonable to treat the two
heavy quarks behave as static color  sources and thus  as a diquark, see e.g.~\cite{Brodsky:2011zs,Yan:2018zdt}.
The effective distance between the two heavy quarks can be estimated to be $r_{QQ}\sim 1/{m_Qv}$, where
$v$ is the four-velocity of the baryon. Further, the distance between one of the  heavy quarks and the
light quark is approximately $r_{Qq}\sim 1/{\Lambda_{QCD}}$.
Since $r_{QQ}/r_{Qq}\sim 1/m_{Q} \ll 1$, the two heavy quarks can be combined to be a point-like diquark.
In the heavy diquark limit,
the heavy diquark system can be treated as a static color source in the $\bar 3$ representation,
just like a heavy anti-quark. Some earlier papers
\cite{Georgi:1990ak,Carone:1990pv,Flynn:2007qt,Nguyen:1993dw} have  used the  heavy quark-diquark symmetry to
simplify  the transition form factors.

In this work, we will try to develop a heavy diquark effective theory (HDiET), whose Lagrangian is
expanded in  powers of $r_{QQ}/r_{Qq}$. At leading-order, the diquark appears as a point-like scalar or
axial-vector particle described by a scalar or axial-vector field in
the color $\bar 3$ representation. The scalar HDiET has been developed in \cite{An:2018cln}, where the
leading order (LO) effective Lagrangian
coupling two scalar diquarks and two light quarks was obtained. In this work, we will first construct
HDiET for both  scalar and axial-vector diquarks.  For the transition form factors we will assume the
applicability of HDiET, and  by assuming the diquark to be a point-like particle, we can construct the
weak and electromagnetic transition currents of the diquarks according to the $\rm SU(2)$ heavy flavor
symmetry and $\rm U(1)$ symmetry. On the other hand,  in the large recoil region, the diquark currents
will be derived through the matching between QCD and HDiET at tree level. We then show that
the six transition  form factors of doubly heavy baryon semi-leptonic decay can be reduced into only one
soft function. Furthermore, it will be shown that this soft function is an universal quantity which
is nothing but the well known Isgur-Wise function in HQET for heavy meson decays.  These results  can
be used in the phenomenology studies.

This article is organized as follows: In section~II, we construct the LO diquark effective theory (DiET)
Lagrangian including the kinetic part as well as the terms coupling with weak and electromagnetic fields.
The DiET is also transformed to HDiET in the heavy diquark limit.
In section~III, we derive the diquark transition currents both from symmetry and tree level matching.
Section~IV focuses on the semi-leptonic decays of doubly heavy baryons. We perform a reduction of the
transition matrix element, where a universal soft function is factorized out and the $q^2$ distributions of
all the six form factors are completely determined from it. The resulting form factors are used to
predict the semi-leptonic decay widths. Section~V contains our conclusions.

\section{Heavy Diquark Effective Theory}
\label{sec:DiquarkEFT}

\subsection{Effective Lagrangian for Scalar and Axial-vector   Diquark}
In this section we will construct the DiET at leading order. The first step is to write down the diquark
effective Lagrangian. We denote the scalar and axial-vector diquark field as $S^i$ and $X^i_{\mu}$,
where $i$ is the $\bar 3$ color index.
The free scalar diquark Lagrangian is simply
\begin{align}
{\cal L}_S={1\over 2}\partial_{\mu}S^{i\dagger}\partial^{\mu}S^i-{1\over 2}m_X^2S^{i\dagger}S^i.\label{SLagr}
\end{align}
Here we have assumed that both the scalar and axial-vector diquark have the same mass $m_X$. On the other
hand, to construct the axial-vector diquark Lagrangian, one should be aware of that $X^i_{\mu}$ is a
matter field in the color fundamental representation $\bar 3$, instead of the adjoint representation which
belongs to the standard gauge fields. Therefore, the axial-vector diquark field is not required to couple
with any conserved current, and it seems not necessary to construct the effective Lagrangian with the
building blocks of the strength tensor 
$F_{\mu\nu}^i=\partial_{\mu}X^i_{\nu}-\partial_{\nu}X^i_{\mu}$ as is done for Yang-Mills theory. Instead,
one can  write down a general form
\begin{equation}
  {\cal L}_{X}=a\  \partial_{\mu}X^i_{\nu} \partial^{\mu}X^{\nu}_i+b\  \partial_{\mu}X^i_{\nu} \partial^{\nu}X^{\mu}_i
  +c\ X^i_{\mu}X^{\mu}_i\label{primaryLag}~.
\end{equation}
However, note that $X_{\mu}$ has four components while a spin-1 particle has only three physical degrees of
freedom. According to the canonical theory, one needs to introduce two second-class constraints for the
Hamiltonian to remove one redundant canonical
variable as well as its conjugate momentum. As a result, one still arrives at a gauge-field-like
Lagrangian
\begin{align}
{\cal L}_{X} & =-\frac{1}{4}F_{\mu\nu}^{i\dagger}F^{\mu\nu i}+\frac{1}{2}m_{X}^{2}X_{\mu}^{i\dagger}X^{\mu i},\label{DiETLag1}
\end{align}
with an on-shell constraint condition $\partial_{\mu}X^{i\mu}=0$.

Since the diquark is composed of two flavored heavy quarks, it is natural to dress the diquark fields
with certain representation in the flavor space.  Notice that in QCD, heavy quarks include bottom and charm.
If we approximately assume $m_b\sim m_c\to\infty$, the mass matrix for $(b, c)^T$ is almost diagonal so
that there exists a flavor $\rm{SU}(2)$ symmetry for the heavy quark sector of the QCD Lagrangian.
Furthermore, in HQET, the leading power Lagrangian ${\bar Q}_v i v\cdot D Q_v$ is exactly invariant under
the flavor $\rm{SU}(2)$ transformation. Such a transformation on a multiplet  $Q=(b,c)^T$ is denoted as
$Q=(b,c)^T$, $Q\to UQ,\ U\in \rm{SU}(2)$. Besides the $\rm{SU}(2)$ flavor symmetry, there is also a
$\rm U(1)$ symmetry which corresponds to the electromagnetic (EM) interaction,
$Q\to U_c Q,\ U_c\in \rm{U}(1)$, where
\begin{equation}
U_{c}={\rm exp\left[i{\cal Q}\ \theta\right]}\in{\rm U}(1),~~~{\cal Q}=\begin{pmatrix}-1/3&0\\
 0& 2/3
\end{pmatrix}~.
\end{equation}

As an effective theory of QCD, DiET should also reflect the $\rm{SU}(2)\times\rm U(1)$ symmetry. In the
flavor space, a diquark field can be considered to have the structure $q^i q^j$, where $i,\ j=b\ \rm{or} \ c$
are flavor indexes. Thus a diquark field should be represented by a $2\times 2$ matrix
\begin{align}
S=\begin{pmatrix} 0& S_{bc}\\
-S_{bc} &0 
\end{pmatrix},\ \ \ 
X_{\mu}=\begin{pmatrix}X_{bb\mu} & X_{bc\mu}\\
X_{bc\mu} & X_{cc\mu}
\end{pmatrix}~.\label{XandSMatrix}
\end{align}
Note that the representation for a scalar diquark is anti-symmetric while the representation for an
axial-vector diquark is symmetric. Under $\rm{SU}(2)\times\rm U(1)$, they transform as 
\begin{align}
S \to U_{(c)} S U_{(c)}^T,~~~X_{\mu} \to U_{(c)} X_{\mu} U_{(c)}^T,~~~U_{(c)}\in \rm{SU}(2)
\ or\  \rm{U}(1)~.
\label{SU2forXS}
\end{align}
With these matrixes as basic building blocks, one can construct a $\rm{SU}(2)\times\rm U(1)$ invariant
diquark Lagrangian.  An efficient way to realize these symmetries is to apply the spinor representation
for the diquark fields. Following Ref.~\cite{Burdman:1992gh}, one firstly combines the spin-1 and spin-0
diquarks to be a multiplet, which is described by a bilinear spinor field $\Sigma$
\begin{align}
\Sigma=(X_{\mu}\gamma^{\mu}+S\gamma_{5})C,
\end{align}
where  $C$ is the charge conjugating matrix. A reason to choose such form  is due to the Lorentz covariance.
Under a general Lorentz transformation $X^{\mu}\to\Lambda^{\mu}_{\nu}X^{\nu}$, one can show that
$\Sigma$ does transform in the expected manner, $\Sigma\to\Lambda_{1/2}\ \Sigma\ \Lambda^T_{1/2}$.
In addition, in  momentum space the equation of motion of the two constituent heavy quarks yakes the
form $\slashed v_d\ \Sigma=\Sigma\ \slashed v_d^T=\Sigma$. Note that since the diquark is treated as a
point-like particle, both the two constituent heavy quarks and the diquark itself share a common
velocity $v_d$, so that it is reasonable to operate with the same slash $\slashed v_d$ on the both sides
of $\Sigma$. Therefore, we can define  $\Sigma^{\prime}$
\begin{align}
\Sigma^{\prime}(v_d)=\frac{1+\slashed v_d}{2}\ \Sigma(v_d)\ \frac{1+\slashed v_d^T}{2}
= \frac{1+\slashed v_d}{2}\left[X_{\mu}(v_d)\gamma^{\mu}+S(v_d)\gamma_{5}\right]C=\frac{1+\slashed v_d}{2}\
\Sigma(v_d)~.
\end{align}
To obtain the second equality we have used the on-shell constraint $v_{d}\cdot X(v_d)=0$. After transforming
$\Sigma^{\prime}(v_d)$ into  coordinates space, we can define a multiplet field $K(x)$ as
\begin{align}
K(x)=\frac{i\slashed \partial+m_X}{2m_X}\ \Sigma(x),~~~~\bar K(x)=\gamma^0 K^{\dagger}(x)
\gamma^0=\bar \Sigma(x) \frac{-i\overleftarrow{\slashed\partial}+m_X}{2m_X},
\end{align}
where $\bar \Sigma(x)=\gamma^0 \Sigma^{\dagger}(x) \gamma^0$. According to Eq.~(\ref{SU2forXS}), under
$\rm{SU}(2)\times\rm U(1)$ transformation, $K$ and $\bar K$ transform in the same manner as $S, X^{\mu}$ and
$S^{\dagger}, X^{\mu\dagger}$.
Therefore the kinematic Lagrangian of DiET is just the simplest globally $\rm{SU}(2)\times\rm U(1)$ invariant
Lagrangian constructed by $K$, $\bar K$, $m_X$ and one derivative operator
\begin{align}
{\cal L}_{DiET}^{kin}=\frac{1}{2}m_X {\rm Tr}\left[\bar K (i \slashed \partial -m_X) K\right]~,
\end{align}
where the trace acts in both flavor and spinor spaces. After expressing this equation in terms
of $X_{\mu}$ and $S$, the kinematic Lagrangian takes the form of a combination of a spin-1 part and  a spin-0 part
\begin{align}
{\cal L}_{DiET}^{kin}=-{1\over 2}{\rm Tr}^f\left[\partial_{\nu}X_{\mu}^{\dagger}\partial^{\nu}X^{\mu}-m_X^2X_{\mu}^{\dagger}X^{\mu}\right]+{1\over 2}{\rm Tr}^f\left[\partial_{\mu}S^{\dagger}\partial^{\mu}S-m_X^2S^{\dagger}S\right],
\end{align}
where ${\rm Tr}^f$ only acts in the flavor space. Compared with Eq.~(\ref{DiETLag1}), this equation
has no $\partial_{\mu}X_{\nu}^{\dagger}\partial^{\nu}X^{\mu}$ term. The reason is that in the heavy quark limit,
the diquark field is a very massive field, which is approximately on shell and satisfies the constraint
$\partial_{\mu}X^{\mu}=0$.

Next, let us  consider how the diquark field couples to external sources. At the quark level,
the weak and the EM coupling come from the coupling terms in QCD
\begin{align}
{\cal L}_{QCD}^{cou}=\bar Q_j \left[V^{\mu}\gamma_{\mu}(1-\gamma_5)+A^{\mu}\gamma_{\mu}\right]_{ji} Q_i
= {\rm Tr}\left[Q\bar Q J\right],\label{QCDcoup}
\end{align}
where $i, j=b\ {\rm or} \ c$ are flavor indexes, $V_{\mu}=V_{\mu}^{a}T^{a}$, $A_{\mu}=A^{em}_{\mu}\cal Q$ and
$J_{ij}=V_{ij}^{\mu}\gamma_{\mu}(1-\gamma_5)+A_{ij}^{\mu}\gamma_{\mu}=L_{ij}+A_{ij}$. The trace acts in both
flavor and spinor spaces. Note that this coupling term is invariant under $\rm{SU}(2)\times\rm U(1)$
transformations if $J$ is assumed to transform as $J\to U_{(c)} J U_{(c)}^{\dagger}$. Therefore, at the
diquark level, the simplest global $\rm{SU}(2)\times\rm U(1)$ invariant coupling terms with 
external source $J$ transfroming in this way are
\begin{align}
{\cal L}_{DiET}^{cou}=\frac{\lambda_1}{2} m_X {\rm Tr}[\bar K J K]+\frac{\lambda_2}{2} m_X {\rm Tr}[\bar K K J^T]~.
\end{align}
Here, $\lambda_1$ and $\lambda_2$ are two independent coupling constants.  After being expressed in terms of
$X_{\mu}$ and $S$, the coupling Lagrangians of the X-J-X, S-J-X, X-J-S and S-J-S types are given by
\begin{align}
{\cal L}_{XJX} & =-i\ {\rm Tr}^{f}\left[F_{\nu\mu}^{\dagger}\langle J^{\mu},X^{\nu}\rangle_{+}-X_{\nu}^{\dagger}\langle
J_{\mu},F^{\nu\mu}\rangle_{+}+i\ \tilde{F}_{\mu\nu}^{\dagger}\langle V^{\mu},X^{\nu}\rangle_{+}-i\ X_{\nu}^{\dagger}
\langle V_{\mu},\tilde{F}^{\nu\mu}\rangle_{+}\right],\label{LagrXJX}\\
{\cal L}_{SJX} & =\frac{1}{m_{X}}{\rm Tr}^{f}\left[\partial_{\nu}S^{\dagger}\langle V_{\mu},F^{\nu\mu}\rangle_{-}
+i\ \partial_{\nu}S^{\dagger}\langle J_{\mu},\tilde{F}^{\nu\mu}\rangle_{-}+m_{X}^{2}S^{\dagger}\langle V_{\mu},X^{\mu}
\rangle_{-}\right],\\
{\cal L}_{XJS} & =\frac{1}{m_{X}}{\rm Tr}^{f}\left[F_{\nu\mu}^{\dagger}\langle V^{\mu},\partial^{\nu}S\rangle_{-}
-i\ \tilde{F}_{\mu\nu}^{\dagger}\langle J^{\mu},\partial^{\nu}S\rangle_{-}+m_{X}^{2}X^{\dagger\mu}\langle V_{\mu},
S\rangle_{-}\right],\\
{\cal L}_{SJS} & =-i\ {\rm Tr}^{f}\left[\partial_{\mu}S^{\dagger}\langle J^{\mu},S\rangle_{+}-S^{\dagger}\langle
J^{\mu},\partial_{\mu}S\rangle_{+}\right],\\
{\cal L}_{DiET}^{cou}&={\cal L}_{XJX} +{\cal L}_{SJX} +{\cal L}_{XJS} +{\cal L}_{SJS}~, \label{LagrSJS}
\end{align}
where ${\rm Tr}^{f}$ only acts in the flavor space and $J_{\mu}=V_{\mu}+A_{\mu}$. ${\tilde F}_{\mu\nu}=
\frac{1}{2}\epsilon_{\mu\nu\alpha\beta}F^{\alpha\beta}$ is the dual field strength tensor. We have also defined
two kinds of commutators in the flavor space
\begin{equation}
\langle A,B\rangle_{\pm}=\frac{\lambda_1}{2}A\ B\pm\frac{\lambda_2}{2}B\ A^{T}~.
\end{equation}

\subsection{Heavy Diquark Effective Theory (HDiET)}
A diquark in the color $\bar 3$ representation interacts with gluons in a similar way as a an anti-quark.
Replacing the ordinary derivatives in Eq.~(\ref{SLagr}) and Eq.~(\ref{DiETLag1}) with covariant derivatives,
one can introduce the coupling of a diquark and a gluon 
\begin{align}
{\cal L}_S&={1\over 2}(D_{\mu}S)^{i\dagger}(D^{\mu}S)^i-{1\over 2}m_X^2S^{i\dagger}S^i,\label{DiETLagS}\\
{\cal L}_{X} &=-\frac{1}{2}\big[(D_{\mu}X_{\nu})^{i\dagger}(D^{\mu}X^{\nu})^i-(D_{\mu}X_{\nu})^{i\dagger}(D^{\nu}X^{\mu})^i\big]+\frac{1}{2}m_{X}^{2}X_{\mu}^{i\dagger}X^{\mu i}\label{DiETLag},
\end{align}
where $D_{\mu}=\partial_{\mu}-i g_d A^a_\mu{\bar t}^a$, $g_d$ is the effective coupling constant between the
diquark and the gluon. In the heavy diquark limit, to expand the Lagrangian in power of $1/m_{X}^{2}$, one has
to separate the diquark field into a static part and a residual part as is done in with the heavy quark in HQET.

For the case of scalar diquark, the $1/m_{X}^{2}$ expansion is trivial. By factorizing out an exponential
phase $S={\rm exp}[-im_X v\cdot x]S_v$, with $v$ the four velocity of the baryon, Eq.~(\ref{DiETLagS}) becomes
\begin{align}
{\cal L}_S=im_X S^{\dagger}_v v\cdot D S_v-\frac{1}{2}S^{\dagger}_v D^{2}S_v~.
\label{HSLagr}
\end{align}
Note that each covariant derivative scales as $\Lambda_{QCD}$. Thus in the heavy diquark limit, the second
term in Eq.~(\ref{HSLagr}) is suppressed by $\Lambda_{QCD}/m_X$ compared with the first term.
Furthermore, at the leading order, $S_v$ is massless and its propagator is simply
\begin{align}
D_S(k)=\frac{i}{m_X v\cdot k}~.
\end{align}

In case of an axial-vector diquark, just factorizing out an exponential phase is not enough. In the heavy
diquark limit, one has to separate $X^{\mu}$ into a static part ${\rm exp}[-im_X v\cdot x]X_v^{\mu}$ which
satisfies $v\cdot X_v=0$ instead of $v_d\cdot X_v=0$, as well as a residual part ${\rm exp}[-im_X v\cdot x]
Y_v^{\mu}$, which is suppressed as $Y_v^{\mu}\sim (\Lambda_{QCD}/m_X)X_v^{\mu}$.
Also note that both $X_v^{\mu}$ and $Y_v^{\mu}$ are dominated by the small momentum $k\sim\Lambda_{QCD}$.
Let us introduce two projection operators $P^{\mu}_{~\nu}$ and $T^{\mu}_{~\nu}$,
\begin{align}
P^{\mu}_{~\nu}=\delta^{\mu}_{~\nu}-v^{\mu}v_{\nu},&~~~T^{\mu}_{~\nu}=v^{\mu}v_{\nu},\\
P^{\mu}_{~\nu}+T^{\mu}_{~\nu}=\delta^{\mu}_{~\nu},~~~P^{\mu}_{~\nu}P^{\nu}_{~\sigma}=P^{\mu}_{~\sigma},
~~~&T^{\mu}_{~\nu}T^{\nu}_{~\sigma}=T^{\mu}_{~\sigma},~~~P^{\mu}_{~\nu}T^{\nu}_{~\sigma}=v_{\mu}P^{\mu}_{~\nu}=0.\nonumber
\end{align}
Using the projection operators, one can project out the static part $X_v^{\mu}$ and the residual part
$Y_v^{\mu}$ of the heavy axial-vector diquark field $X^{\mu}$
\begin{align}
X_v^{\mu}=e^{im_X v\cdot x}\ P^{\mu}_{~\nu}X^{\nu},~~~~Y_v^{\mu}=e^{im_X v\cdot x}\ T^{\mu}_{~\nu}X^{\nu}~,
\end{align}
which satisfy $v\cdot X_v=0~\text{and}~(\cdots X_v^{\mu})^{\dagger}(\cdots Y_{v\mu})=(\cdots Y_v^{\mu})^{\dagger}
(\cdots X_{v\mu})=0$, where the dots represent any possible insertion of covariant derivatives.  Then
the full diquark field can be separated as
\begin{equation}
X^{\mu}=e^{-im_X v\cdot x}(X_v^{\mu}+Y_v^{\mu})~.
\label{Xsepar}
\end{equation}
Inserting Eq.~(\ref{Xsepar}) into Eq.~(\ref{DiETLag}), and using integration by part $\overleftarrow{D}=-D$
to make all the covariant derivatives act on the $X, Y$ fields instead of the $X^{\dagger}, Y^{\dagger}$ fields,
one finally arrives at
\begin{align}
{\cal L}_{X} =&-im_{X}X_{v\mu}^{\dagger}v\cdot D X_{v}^{\mu}+\frac{1}{2}X_{v\mu}^{\dagger}(D^{2}g^{\mu\nu}-D^{\nu}D^{\mu})X_{v\nu}\nonumber \\
 & -im_{X}Y_{v\mu}^{\dagger}\left[g^{\mu\nu}(v\cdot D)-\frac{1}{2}(v^{\nu}D^{\mu}+v^{\mu}D^{\nu}-im_{X}v^{\mu}v^{\nu})+\frac{i}{2m_{X}}(D^{2}g^{\mu\nu}-D^{\nu}D^{\mu})\right]Y_{v\nu}\nonumber \\
& +\frac{i}{2}m_{X}X_{v\nu}^{\dagger}\left(v^{\mu}D^{\nu}+\frac{i}{m_{X}}D^{\mu}D^{\nu}\right)Y_{v\mu}+\frac{i}{2}m_{X}Y_{v\nu}^{\dagger}\left(v^{\nu}D^{\mu}+\frac{i}{m_{X}}D^{\mu}D^{\nu}\right)X_{v\mu}~.
\label{LagrXandY}
\end{align}
From the Lagrangian Eq.~(\ref{LagrXandY}), one finds that $X_v^{\mu}$ is a massless field, while $Y_v^{\mu}$ is
massive due to the non-diagonal mass term $-(m_X^2/2)v^{\mu}v^{\nu}Y_{v\mu}^{\dagger}Y_{v\nu}$. To obtain
an effective theory containing only the massless field $X_v^{\mu}$, one needs to integrate out the heavy
degree of freedom $Y_v^{\mu}$. One way to realize this is to use the saddle point approximation,
where one first solves the equation of motion of the heavier field $Y_v^{\mu}$ while keeping $X_v^{\mu}$ fixed.
The solution is
\begin{align}
\left[g^{\mu\nu}(v\cdot D)-\frac{1}{2}(v^{\nu}D^{\mu}+v^{\mu}D^{\nu}-im_{X}v^{\mu}v^{\nu})+\frac{i}{2m_{X}}(D^{2}g^{\mu\nu}-D^{\nu}D^{\mu})\right]Y_{v\nu}\nonumber\\
=\frac{1}{2}\left(v^{\mu}D^{\nu}+\frac{i}{m_{X}}D^{\nu}D^{\mu}\right)X_{v\nu}~.
\label{EOMofY}
\end{align}
It is not simple to solve this matrix equation directly. To simplify it, we can multiply with $v_{\mu}$ on
both sides of the equation 
\begin{align}
\left[iv^{\mu}+\frac{v^{\mu}(v\cdot D)-D^{\mu}}{m_{X}}+i\frac{D^{2}v^{\mu}-D^{\mu}(v\cdot D)}{m_{X}^{2}}\right]
Y_{v\mu}=\left[\frac{D^{\mu}}{m_{X}}+i\frac{D^{\mu}(v\cdot D)}{m_{X}^{2}}\right]X_{v\mu}, \label{simEOMofY}
\end{align}
and introduce a power counting scheme to solve this equation perturbatively. Note that each covariant
derivative $D$ scales as $\Lambda_{QCD}$ which is small compared to $m_X$. So by counting the number
of ${\kappa}=D/{m_X}$, we can conclude that
\begin{equation}
v^{\mu}\sim{\cal O}(1);\ \ \ \frac{v^{\mu}(v\cdot D)-D^{\mu}}{m_{X}},\ \frac{D^{\mu}}{m_{X}}\sim{\cal O}(\kappa);
\ \ \ \frac{D^{2}v^{\mu}-D^{\mu}(v\cdot D)}{m_{X}^{2}},\ \frac{D^{\mu}(v\cdot D)}{m_{X}^{2}}\sim{\cal O}(\kappa^2)~.
\end{equation}
Since $Y_v^{\mu}$ is orthogonal to $X_v^{\mu}$, $Y_v^{\mu}$ cannot involve a term like $\text{const}\times X_v^{\mu}$. 
The solution of Eq.~(\ref{simEOMofY}) up to ${\cal O}(\kappa^3)$ is given as:
\begin{equation}
Y_v^{\mu}= -\frac{i}{m_X}v^{\mu}D_{\nu}X_v^{\nu}+\frac{1}{m_X^2} D_{\nu}D^{\mu}X_v^{\nu}
+{\cal O}(\kappa^3)~.
\label{SolutionY}
\end{equation}
After inserting this solution of $Y_v^{\mu}$ back to Eq.~(\ref{LagrXandY}), one finally obtains the effective
Lagrangian in the form of a power expansion
\begin{align}
{\cal L}_{X} =&-im_{X}X_{v\mu}^{\dagger} v\cdot D X_{v}^{\mu}+\frac{1}{2}X_{v\mu}^{\dagger}D^{2}X_{v}^{\mu}+\frac{i}{2}g_d X_{v\mu}^{\dagger}{\bar G}^{\mu\nu}X_{v\nu}\nonumber\\
&+\frac{i}{2 m_X}X_{v\mu}^{\dagger}\left\{D^{\mu}D^{\nu},v\cdot D\right\}X_{v\nu}+{\cal O}\left(1/m_X^2\right)~,
\label{HDiETLagr}
\end{align}
where ${\bar G}_{\mu\nu}=G^a_{\mu\nu}{\bar t}^a$ is the gluon tensor. In the Eq.~(\ref{HDiETLagr}), the
second term represents the heavy diquark kinetic energy while the third term corresponds to the chromomagnetic
coupling. These two terms are consistent with those given in Ref.~\cite{Savage:1990di,Fleming:2005pd,Hu:2005gf},
where a non-relativistic approach is used.  The propagator of the massless heavy axial-vector diquark is
\begin{align}
D_X^{\mu\nu}(k)=\frac{-i}{m_X v\cdot k}(g^{\mu\nu}-v^{\mu}v^{\nu})~.
\label{FreeDiProp}
\end{align}

The heavy diquark can only couple to soft gluons. Through the following field redifinition, one can
decouple the diquark field from gluon field:
\begin{align}
X_{v\mu}=P\Big\{{\rm exp}\Big[ig\int_{-\infty}^{v\cdot x}ds\ v\cdot A(s)\Big]\Big\}\tilde{X_v}_{\mu}
& = W\Big[\begin{array}{c}
x\\
v
\end{array}\Big]{\tilde X}_{v\mu},~~~
S_{v}=W\Big[\begin{array}{c}
x\\
v
\end{array}\Big]{\tilde S}_{v}.\label{decoTrans}\\
(v\cdot D)X_v^{\mu}=W\Big[\begin{array}{c}
x\\
v
\end{array}\Big](v\cdot\partial){\tilde X}_v^{\mu},\ \ \ \ \ &(v\cdot D)S_v=W\Big[\begin{array}{c}
x\\
v
\end{array}\Big](v\cdot\partial){\tilde S}_v~.
\end{align}
Using the decoupling transformation, one can replace all the covariant derivatives in Eq.~(\ref{HDiETLagr})
by ordinary derivatives, while the $X$ field should be replaced by the dressed field $\tilde X$.

\section{Heavy to heavy Baryonic Transitions }
\subsection{Diquark Transition Currents from Symmetry}

When using DiET to study doubly heavy baryon decays ${\cal B}_{bQ}\to{\cal B}_{cQ} l\nu$, for instance when the
$bb$ diquark turns into the $bc$ diquark through the  $V-A$ current $\bar{c}\gamma_{\mu}(1-\gamma_5)b$, or
electromagnetic transitions ${\cal B}_{Q_1Q_2}\to{\cal B}_{Q_1Q_2} \gamma^{*}$ induced by the vector current
$\bar{Q}\gamma_{\mu}Q$, one needs to express the corresponding currents in terms of the diquark fields instead of
the heavy quark fields. Particularly, if we approximate the diquark as a point like particle, we require
the four most general kinds of diquark currents 
\begin{align}
X_{\alpha}^{\dagger}\Gamma_{\mu}^{\alpha\beta}[\overleftarrow{\partial},\partial]X_{\beta},
~~~S^{\dagger}\Gamma_{\mu}^{\beta}[\overleftarrow{\partial},\partial]X_{\beta},
~~~X_{\beta}^{\dagger}\Gamma_{\mu}^{\beta}[\overleftarrow{\partial},\partial]S~~~\text{and}
~~~S^{\dagger}\Gamma_{\mu}[\overleftarrow{\partial},\partial]S~,
\label{DiVecCurr1}
\end{align}
which correspond to pure axial-vector, axial-vector to scalar, scalar to axial-vector and pure scalar transitions.
Note that $\Gamma_{\mu}^{\alpha\beta}, \Gamma_{\mu}^{\beta}$ and $\Gamma_{\mu}$ depend on the momentum of the initial
and final diquarks. In the heavy diquark limit we can simply replace the $\overleftarrow{\partial},\ \partial$
with the four-velocities of the final and initial baryons $i v_2,\ -i v_1$, with $w=v_1\cdot v_2~ \text{close to}~ 1$
for the low recoil region. 

Consider first  the case of $V-A$ weak current $\bar{c}\gamma_{\mu}(1-\gamma_5)b$. According to Eq.~(\ref{QCDcoup}),
it is just a current coupling to the external source $V_{\mu}^{1}+iV_{\mu}^{2}$, which can be found from the expansion
\begin{align}
{\rm Tr}\left[Q\bar Q L\right]=\ &\bar{c}\gamma^{\mu}(1-\gamma_5)b\ (V_{\mu}^{1}+iV_{\mu}^{2})+\bar{b}\gamma^{\mu}
(1-\gamma_5)c\ (V_{\mu}^{1}
-iV_{\mu}^{2})\nonumber\\
&+\left[\bar{b}\gamma^{\mu}(1-\gamma_5)b-\bar{c}\gamma^{\mu}(1-\gamma_5)c\right]\ V_{\mu}^{3}.
\end{align}
Straightforwardly, one can conclude that the $\bar{c}\gamma_{\mu}(1-\gamma_5)b$ current can be produced by
operating with a derivative on the part of the Lagrangian of QCD that contains the couplings to the external
fields
\begin{align}
\bar{c}\gamma_{\mu}(1-\gamma_5)b=\frac{\partial}{\partial(V^{\mu}_{1}+iV^{\mu}_{2})}{\cal L}^{\rm coup}_{QCD}~.
\end{align}
On the other hand, on the diquark level,  if one performs the same derivative operation on the DiET
Lagrangian Eq.~(\ref{LagrXJX}-\ref{LagrSJS}), one arrives at the $V-A$ currents in the DiET form
\begin{align}
J_{\mu}^{Transition}= \frac{\partial}{\partial(V^{\mu}_{1}+iV^{\mu}_{2})}\left[{\cal L}_{XJX}+{\cal L}_{SJX}
  +{\cal L}_{XJS}+{\cal L}_{SJS}\right]~.
\label{CurrDeriv}
\end{align}
Explicitly for $X\to X$, $X\to S$ and $S\to S$ transitions, one has
\begin{align}
J_{\mu}^{X\to X}  =& -\frac{1}{2}(\lambda_{1}+\lambda_{2})\Big[i\left(\partial_{\nu}X_{bc\mu}^{\dagger}X_{bb}^{\nu}-\partial_{\mu}X_{bc\nu}^{\dagger}X_{bb}^{\nu}-X_{bc\nu}^{\dagger}\partial^{\nu}X_{bb\mu}+X_{bc\nu}^{\dagger}\partial_{\mu}X_{bb}^{\nu}\right)\nonumber \\
 & -\epsilon_{\alpha\beta\mu\rho}\left(\partial^{\rho}X_{bc}^{\dagger\alpha}X_{bb}^{\beta}-X_{bc}^{\dagger\alpha}\partial^{\rho}X_{bb}^{\beta}\right)+(bc \to cc)\Big]~,\label{XXCurrSym}\\
 J_{\mu}^{X\to S}  =&-\frac{1}{2m_{X}}(\lambda_{1}+\lambda_{2})\left(\partial_{\nu}S_{bc}^{\dagger}\partial^{\nu}X_{bb\mu}-\partial_{\nu}S_{bc}^{\dagger}\partial_{\mu}X_{bb}^{\nu}+m_{X}^{2}S_{bc}^{\dagger}X_{bb\mu}\right)\nonumber \\
 & +\frac{i}{2m_{X}}(\lambda_{1}+\lambda_{2})\epsilon_{\rho\mu\sigma\beta}\partial^{\rho}S_{bc}^{\dagger}\partial^{\sigma}X_{bb}^{\beta}~,\label{SXCurrSym}\\
 J_{\mu}^{S\to X}= & -\frac{1}{2m_{X}}(\lambda_{1}+\lambda_{2})\left(\partial_{\mu}X_{cc\nu}^{\dagger}\partial^{\nu}S_{bc}-\partial_{\nu}X_{cc\mu}^{\dagger}\partial^{\nu}S_{bc}-m_{X}^{2}X_{cc\mu}^{\dagger}S_{bc}\right)\nonumber\nonumber \\
 & -\frac{i}{2m_{X}}(\lambda_{1}+\lambda_{2})\epsilon_{\alpha\rho\mu\sigma}\partial^{\rho}X_{cc}^{\dagger\alpha}\partial^{\sigma}S_{bc}~.
\label{XSCurrSym}
\end{align}
Note that the antisymmetric $S$ has only one non-vanishing component $S_{bc}$, for flavor changing processes
$b\to c$ there is no $S\to S$ transition. Similarly, the electromagnetic currents $I_{\mu}^{\rm Transition}$ can
be derived by acting with a derivative on $A_{em}^{\mu}$,
\begin{align}
  &I_{\mu}^{\rm Transition}= \frac{\partial}{\partial A_{em}^{\mu}}\left[{\cal L}_{XJX}+{\cal L}_{SJX}
    +{\cal L}_{XJS}+{\cal L}_{SJS}\right],\label{EMCurrDeriv}\\
&I_{\mu}^{X\to X}=-\frac{i}{4}C_{X}(\lambda_1+\lambda_2)\left(\partial^{\dagger\nu}X_{\mu}X_{\nu}-\partial_{\mu}X_{\nu}^{\dagger}X^{\nu}-X_{\nu}^{\dagger}\partial^{\nu}X_{\mu}+X_{\nu}^{\dagger}\partial_{\mu}X^{\nu}\right)~,\label{XXemCurrSym}\\
&I_{\mu}^{X\to S}=-\frac{i}{2m_X}\epsilon_{\rho\mu\alpha\beta}(\lambda_1+\lambda_2)\partial^{\rho}S_{bc}^{\dagger}\partial^{\alpha}X_{bc}^{\beta}~,\label{SXemCurrSym}\\
&I_{\mu}^{S\to V}=-\frac{i}{2m_X}\epsilon_{\alpha\rho\mu\beta}(\lambda_1+\lambda_2)\partial^{\rho}X_{bc}^{\dagger\alpha}\partial^{\beta}S_{bc},\label{XSemCurrSym}\\
&I_{\mu}^{S\to S}=-\frac{i}{6}(\lambda_1+\lambda_2)\left(\partial_{\mu}S_{bc}^{\dagger}S_{bc}-S_{bc}^{\dagger}\partial_{\mu}S_{bc}\right)~,\label{SSemCurrSym}
\end{align}
where $C_X$ is the total electric charge of $X$. It should be mentioned that all the currents in
Eqs.~(\ref{XXCurrSym}-\ref{XSCurrSym}, \ref{XXemCurrSym}-\ref{SSemCurrSym}) are expressed by the full
diquark fields. These expressions are simpler in the heavy diquark limit. According to Eq.~(\ref{Xsepar})
and Eq.~(\ref{SolutionY}), the full diquark fields $X, S$ are related with the effective ones $X_v, S_v$
in HDiET as
\begin{equation}
S=e^{-im_X v\cdot x}S_v,~~~X^{\mu}=e^{-im_X v\cdot x}\left(X_v^{\mu}-\frac{i}{m_X}v^{\mu}D_{\nu}X_{v}^{\nu}\right)~.
\label{XandXv}
\end{equation}
Inserting Eq.~(\ref{XandXv}) into Eq.~(\ref{XXCurrSym}-\ref{XSCurrSym}, \ref{XXemCurrSym}-\ref{SSemCurrSym}),
at leading order, all the derivative operators are simply replaced by the corresponding four velocities
\begin{align}
&J_{\mu}^{X\to X} =\Lambda\  X_{(v)cQ}^{\dagger\alpha}\left[g_{\mu\alpha}v_{2\beta}+v_{1\alpha}g_{\mu\beta}-(v_{1\mu}+v_{2\mu})g_{\alpha\beta}+i\epsilon_{\alpha\beta\mu\rho}(v_{2}^{\rho}+v_{1}^{\rho})\right]X_{(v)bQ}^{\beta}~,\label{HDiETEWCurr1}\\
&J_{\mu}^{X\to S} =-\Lambda\ S_{(v)bc}^{\dagger}\left[(1+w)g_{\mu\beta}-v_{1\mu}v_{2\beta}-i\epsilon_{\rho\mu\sigma\beta}v_{2}^{\rho}v_{1}^{\sigma}\right]X_{(v)bb}^{\beta}~,\label{HDiETEWCurr2}\\
&J_{\mu}^{S\to X} =-\Lambda\ X_{(v)cc}^{\dagger\alpha}\left[-(1+w)g_{\mu\alpha}+v_{2\mu}v_{1\alpha}-i\epsilon_{\rho\mu\sigma\alpha}v_{2}^{\rho}v_{1}^{\sigma}\right]S_{(v)bc}~,\label{HDiETEWCurr3}\\
&I_{\mu}^{X\to X} =\frac{1}{2}C_{X}\Lambda\ X_{(v)}^{\dagger\alpha}\left[g_{\mu\alpha}v_{2\beta}-v_{2\mu}g_{\alpha\beta}+v_{1\alpha}g_{\mu\beta}-v_{1\mu}g_{\alpha\beta}\right]X_{(v)}^{\beta}~,\label{HDiETEMCurr1}\\
&I_{\mu}^{X\to S} =\Lambda\ S_{(v)bc}^{\dagger}\left[-i\epsilon_{\rho\mu\sigma\beta}v_{2}^{\rho}v_{1}^{\sigma}\right]X_{(v)bc}^{\beta},\\
&I_{\mu}^{S\to X} =\Lambda\ X_{(v)bc}^{\dagger\alpha}\left[i\epsilon_{\rho\mu\sigma\alpha}v_{2}^{\rho}v_{1}^{\sigma}\right]S_{(v)bc},\\
&I_{\mu}^{S\to S} =\frac{1}{3}\Lambda\ S_{(v)bc}^{\dagger}\left[v_{2\mu}+v_{1\mu}\right]S_{(v)bc}~,\label{HDiETEMCurr4}
\end{align}
where $\Lambda=(\lambda_{1}+\lambda_{2})m_{X}$. Similarly one can obtain the currents at next-to-leading order
if the second expansion term of $X^{\mu}$ in Eq.~(\ref{XandXv}) is used, but the results will not be
shown explicitly here.

It should be mentioned that like the chromomagnetic coupling in the Eq.~(\ref{HDiETLagr}), one can also introduce
the magnetic couplings of the axial-vector diquark as those given in Ref.~\cite{Hu:2005gf} by NRQCD. Such a
term will contribute an extra EM current suppressed by $1/m_X$ in Eq.~(\ref{HDiETEMCurr1}-\ref{HDiETEMCurr4}).

\subsection{Diquark Transition Currents from Matching}
\begin{figure}
\includegraphics[width=0.9\columnwidth]{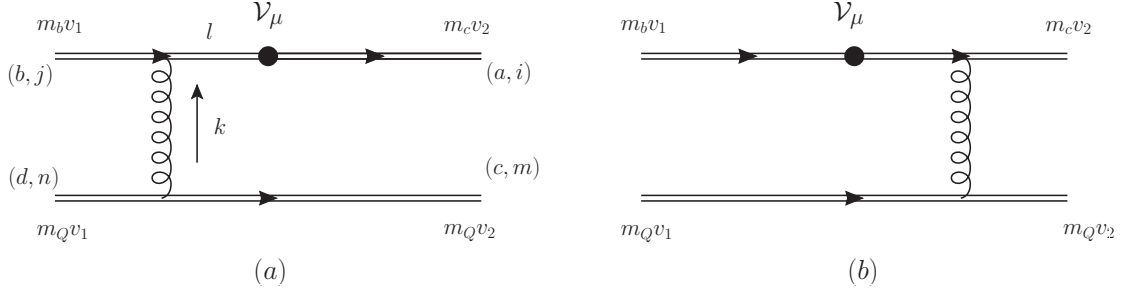} 
\caption{Diquark transition at the quark level. Single gluon exchanges between the two heavy quarks
  are explicitly shown, with the quantum numbers of the quarks are denoted as (spinor index, color index).}
\label{fig:quarklevel} 
\end{figure}

When the recoil is small, to derive the diquark transition currents from symmetries we can assume the
diquark as a point-like particle without any internal structure. Therefore, the currents we get in
Eq.~(\ref{XXCurrSym}-\ref{XSCurrSym}, \ref{XXemCurrSym}-\ref{SSemCurrSym}) are only proportional to
the constant couplings $\lambda_1, \lambda_2$. On the other hand, if the recoil is large, we should
consider finite sized diquarks where the transition is dominated by hard internal gluon exchange which can
be factorized into short distance coefficients. One way to obtain these short distance coefficients is
to perform a matching between DiET and QCD in the large recoil region, where at the quark level one may
factorize out a hard kernel, with its tree level form  shown in Fig.~\ref{fig:quarklevel}. A hard gluon
is exchanged between the two heavy quarks so that the recoil is large, $q^2~\text{close to zero}$,
and ${\cal V}_{\mu}=\gamma_{\mu}\  \text{or}\  \gamma_{\mu}\gamma_{5}$ is the current vertex.

The calculation of the two diagrams in Fig.~\ref{fig:quarklevel} is straightforward. However, although at
tree level we can set the initial and final quarks to be free, the two quark spins are coupled so that
the total spin should match with the corresponding diquark spin. Particularly, to match with a scalar
or axial-vector diquark, the spinor indexes of the two quarks should be symmetrical or anti-symmetrical.
Consider first the $X\to X$ transition. By equating the velocities of the initial and final two quarks to
be $v_1$ and $v_2$ respectively, the amplitude of the two diagrams in Fig.~\ref{fig:quarklevel} reads
\begin{align}
{\cal M}_{\rm QCD} =&-g_{d}^{2}t_{ij}^{A}t_{mn}^{A}\frac{1}{4m_{c}^{3}(1-w)^{2}}\bar{u}_{(c)[i}^{\{a}\bar{u}_{(Q)m]}^{c\}}
\Big\{\left[{\cal V}_{\mu}\left(\slashed v_{1}-\xi_{1}\slashed v_{2}+1-\xi_{1}\right)\gamma^{\nu}\right]_{ab}
(\gamma_{\nu})_{cd}\nonumber \\
& +\big[\gamma^{\nu}\big(\slashed v_{2}-\xi_{2}\slashed v_{1}+1-\xi_{2}\big){\cal V}_{\mu}\big]_{ab}
(\gamma_{\nu})_{cd}\Big\} u_{(b)[j}^{\{b}u_{(Q)n]}^{d\}},
\end{align}
where $a, b, c, d$ are spinor indices, and $i, j, m, n$ are color indices. Further, $\xi_1=m_Q/(m_Q+m_b)$
and $\xi_2=m_Q/(m_Q+m_c)$. 
For the finite-sized diquark, the corresponding weak transition amplitude is
\begin{align}
{\cal M}_{\rm DiET} & =X_{\alpha}^{\dagger}(v_2)\Gamma_{\mu}^{\alpha\beta}\left[v_1,v_2\right]X_{\beta}(v_1)~.
\label{DiETAmp1}
\end{align}
Here, $X^{\dagger}(v_2), X(v_1)$ should be treated as the polarization vectors of the final and initial
diquarks, and $\Gamma_{\mu}^{\alpha\beta}\left[v_1,v_2\right]$ represents the hard kernel.  Explicitly,
the diquark wave function can be composed of two heavy quark spinors as
\begin{align}
S^{i}(v) & =N_S \epsilon^{ijk}Q_{1j\beta}(v)Q_{2k\gamma}(v)[C\gamma_{5}(1+\slashed v)]_{\beta\gamma}~,\nonumber\\
X_{\mu}^{i}(v) & =N_X \epsilon^{ijk}Q_{1}{}_{j\beta}(v)Q_{2}{}_{k\gamma}(v)[C\gamma_{\mu}(1+\slashed v)]_{\beta\gamma}~,
\label{XWVF}\nonumber
\end{align}
where $i, j, k$ and $\beta, \gamma$ are color and spinor indices, respectively, and $N_S, N_X$ are
normalization factors. Inseting Eq.~(\ref{XWVF}) into Eq.~(\ref{DiETAmp1}) and factorizing an independent
color factor $C\ \delta_{l}^{k}$, one arrives at
\begin{align}
{\cal M}_{\rm DiET} =N_{X_c}N_{X_b}\bar{u}_{(c)}^{i\{a}\bar{u}_{(Q)}^{c\}j}
\epsilon_{ijk}[(1+\slashed v_{2})\gamma_{\alpha}C]_{ac}
\left(\Gamma_{\mu}^{\alpha\beta}\times C\ \delta_{l}^{k}\right)[C\gamma_{\beta}(1+\slashed v_{1})]_{bd}
\epsilon^{lmn}u_{(b)m}^{\{b}u_{(Q)n}^{d\}}~.
\end{align}
The tree level matching demands the equivalence of the amplitudes at the quark
and the diquark level ${\cal M}_{\rm QCD}  ={\cal M}_{\rm DiET}$, thus we can determine the hard kernel as
\begin{align}
\Gamma_{\mu}^{\alpha\beta} =&\frac{g_{d}^{2}}{32N_{X_{bQ}}N_{X_{cQ}}m_{Q}^{3}(1-w)^{2}}\Big\{{\rm tr}
\left[\gamma^{\alpha}{\cal V}_{\mu}\left(\slashed v_{1}-\xi_{1}\slashed v_{2}+1-\xi_{1}\right)\gamma^{\beta}\right]
\nonumber \\
& +{\rm tr}\left[\gamma^{\alpha}\left(\slashed v_{2}-\xi_{2}\slashed v_{1}+1-\xi_{2}\right){\cal V}_{\mu}
\gamma^{\beta}\right]\Big\}~,
\end{align}
and the color factor is $C=-1/3$. Similarly, for $X\to S$ and $S\to X$ transitions, we have
\begin{align}
\Gamma_{\mu}^{\beta}[X\to S] =&\frac{g_{d}^{2}}{32N_{X_{bQ}}N_{S_{cQ}}m_{Q}^{3}(1-w)^{2}}\Big\{{\rm tr}\left[\gamma_{5}
{\cal V}_{\mu}\left(\slashed v_{1}-\xi_{1}\slashed v_{2}+1-\xi_{1}\right)\gamma^{\beta}\right]\nonumber \\
& +{\rm tr}\left[\gamma_{5}\left(\slashed v_{2}-\xi_{2}\slashed v_{1}+1-\xi_{2}\right){\cal V}_{\mu}
\gamma^{\beta}\right]\Big\}~,\\
\Gamma_{\mu}^{\alpha}[S\to X] =&\frac{g_{d}^{2}}{32N_{S_{bQ}}N_{X_{cQ}}m_{Q}^{3}(1-w)^{2}}\Big\{{\rm tr}
\left[\gamma^{\alpha}{\cal V}_{\mu}\left(\slashed v_{1}-\xi_{1}\slashed v_{2}+1-\xi_{1}\right)\gamma_{5}
\right]\nonumber \\
& +{\rm tr}\left[\gamma^{\alpha}\left(\slashed v_{2}-\xi_{2}\slashed v_{1}+1-\xi_{2}\right){\cal V}_{\mu}
\gamma_{5}\right]\Big\}~.
\end{align}
Particularly, for the $V-A$ currents, where ${\cal V}_{\mu}=\gamma_{\mu}\  \text{or}\  \gamma_{\mu}\gamma_{5}$,
the hard kernels are 
\begin{align}
\Gamma_{\mu(V)}^{\alpha\beta}&=  -\frac{g_{d}^{2}}{8N_{X_{bQ}}N_{X_{cQ}}m_{Q}^{3}(1-w)^{2}}\Big[(\xi_{1}+1)v_{2}^{\beta}
g_{\mu}^{\alpha}+(\xi_{1}-1)v_{2\mu}g^{\alpha\beta}-(\xi_{1}+1)v_{2}^{\alpha}g_{\mu}^{\beta}\nonumber \\
&~~~~ -(\xi_{2}+1)v_{1}^{\beta}g_{\mu}^{\alpha}+(\xi_{2}-1)v_{1\mu}g^{\alpha\beta}+(\xi_{2}+1)v_{1}^{\alpha}
g_{\mu}^{\beta}\Big]~,\label{matchXVX}\\
\Gamma_{\mu(A)}^{\alpha\beta}&=-\frac{g_{d}^{2}}{8N_{X_{bQ}}N_{X_{cQ}}m_{Q}^{3}(1-w)^{2}}i
\epsilon_{~~\mu\sigma}^{\alpha\beta}\left[(\xi_{1}-1)v_{2}^{\sigma}+(\xi_{2}-1)v_{1}^{\sigma}\right]~,\\
\Gamma_{\mu(V)}^{\beta}&[X\to S]= \Gamma_{\mu(V)}^{\beta}[S\to X]=0~,\\
\Gamma_{\mu(A)}^{\beta}&[X\to S]= -\frac{g_{d}^{2}}{8N_{X_{bQ}}N_{S_{cQ}}m_{Q}^{3}(1-w)^{2}}\left(2-\xi_1-\xi_2\right)
g_{\mu}^\beta~,\\
\Gamma_{\mu(A)}^{\beta}&[S\to X]= \frac{g_{d}^{2}}{8N_{S_{bQ}}N_{X_{cQ}}m_{Q}^{3}(1-w)^{2}}\left(2-\xi_1-\xi_2\right)
g_{\mu}^\alpha~.
\end{align}
For the  EM currents, the $X\to X$, $X\to S$ and $S\to X$ currents have the same hard kernel as those of the
$V-A$ currents except  for the replacements $m_b\to m_{Q^{\prime}}, m_c\to m_{Q^{\prime}}$. However, the $S\to S$ EM
current is
\begin{align}
\Gamma_{\mu(\rm EM)}[S\to S]= -\frac{g_{d}^{2}}{8N_{S_{bQ}}N_{S_{cQ}}m_{Q}^{3}(1-w)^{2}}\left[(\xi_1-1)v_{2\mu}+(\xi_2-1)
  v_{1\mu}\right]~.
\label{matchSEMS}
\end{align}
Note that the structures shown in Eq.~(\ref{matchXVX}-\ref{matchSEMS}) are different from those in Eq.~(\ref{XXCurrSym}-\ref{SSemCurrSym}). Such differences can be understood because the singular point $w=1$ appearing in the
Eq.~(\ref{matchXVX}-\ref{matchSEMS}) implies that they are only valid in the large recoil region $w\to w_{max}$.

\section{Semi-Leptonic Decays of Doubly Heavy Baryons}
\subsection{Interpolating Fields}

In this section we will focus on semi-leptonic decays of doubly heavy baryons, ${\cal B}_{bQ}\to
{\cal B}_{cQ}\ell\bar\nu$. The transition matrix element of the doubly heavy baryon can be calculated
by the reduction formula
\begin{align}
\langle{\cal B}_{cQ}(P_c)|J_{\mu}(0)|{\cal B}_{bQ}(P_b)\rangle= L(P_{b},P_{c})\int d^{4}xd^{4}y\ e^{iP_{c}\cdot x}
e^{-iP_{b}\cdot y}\langle0|\mathrm{T}\Phi_{cQ}(x)J_{\mu}(0)\Phi_{bQ}^{\dagger}(y)|0\rangle~,
\label{LSZformula}
\end{align}
where $J_{\mu}$ is the current inducing the weak decay. $L(P_{b},P_{c})$ is the operator to pick  out the
initial and final mass pole residues
\begin{equation}
L(P_{b},P_{c})=\lim_{P_{b}^{2}\to M_{b}^{2}}(P_{b}^{2}-M_{b}^{2})\lim_{P_{c}^{2}\to M_{c}^{2}}(P_{c}^{2}-M_{c}^{2})~.
\end{equation}
$\Phi_{cQ}(x)$ and $\Phi_{bQ}(x)$ are the interpolating fields of the final and initial baryon.
Eq.~(\ref{LSZformula}) can be expressed both at the quark level and the diquark level. At the quark level,
$J_{\mu}=\bar c\gamma_{\mu}(1-\gamma_5)b$, and
\begin{align}
\Phi_{Q_{1}Q_{2}}(x) & =N_{Q_1Q_2}\epsilon^{ijk}\bar{\chi}^{\alpha\beta\gamma}q_{i\alpha}(x)Q_{1}{}_{j\beta}(x)
Q_{2}{}_{k\gamma}(x)~,\nonumber\\
\Phi_{Q_{1}Q_{2}}^{\dagger}(x) & =N_{Q_1Q_2}\epsilon_{ijk}\chi_{\alpha\beta\gamma}\bar{Q}_{2}^{i\gamma}(x)\bar{Q}_{1}^{j\beta}(x)
\bar{q}^{k\alpha}(x)~,
\end{align}
where $\chi$ are the Bargmann-Wigner wave functions \cite{Hussain:1994gj}, where the total spin contributed by
the two heavy quarks is $j$. For a spin-1/2 doubly heavy baryon with $j=0$ or $j=1$, and a spin-3/2 baryon
with $j=1$, they are
\begin{align}
&\chi^{1/2(0)}_{\alpha\beta\gamma} =\chi^{1/2(0)}_{\alpha\{\beta\gamma\}}=\frac{1}{2}u_{\alpha}[(1+\slashed v)
\gamma_{5}C]_{\beta\gamma}~,\nonumber\\
&\chi^{1/2(1)}_{\alpha\beta\gamma} =\chi^{1/2(1)}_{\alpha\{\beta\gamma\}}=\frac{1}{2}[(\gamma^{\mu}+v^{\mu})
\gamma_{5}u]_{\alpha}[(1+\slashed v)\gamma_{\mu}C]_{\beta\gamma},\nonumber\\
&\chi^{3/2}_{\alpha\beta\gamma} =\chi^{3/2}_{\{\alpha\beta\gamma\}}=u^{\mu}_{\alpha}[(1+\slashed v)\gamma_{\mu}
C]_{\beta\gamma}~.
\end{align}
The symmetry indices $\beta,\ \gamma$ project out the spin-1 configuration of the two heavy quarks. The conjugate
forms are defined as $\bar{\chi}^{\alpha\beta\gamma}=(\gamma_{0})^{\alpha\alpha^{\prime}}(\gamma_{0})^{\beta\beta^{\prime}}
(\gamma_{0})^{\gamma\gamma^{\prime}}\chi_{\alpha\beta\gamma}$. $\chi_{\alpha\beta\gamma}$ satisfies
\begin{align}
(\slashed v-1)_{\alpha}^{\alpha^{\prime}}\chi_{\alpha^{\prime}\beta\gamma}&=(\slashed v-1)_{\beta}^{\beta^{\prime}}
\chi_{\alpha\beta^{\prime}\gamma}=(\slashed v-1)_{\gamma}^{\gamma^{\prime}}\chi_{\alpha\beta\gamma^{\prime}}=0~,\nonumber\\
&\chi_{\alpha\{\beta\gamma\}}^{1/2}+\chi_{\beta\{\gamma\alpha\}}^{1/2}+\chi_{\gamma\{\alpha\beta\}}^{1/2}=0~.
\end{align}
On the other hand, we can equivalently express Eq.~(\ref{LSZformula}) at diquark level, with the assumption
that the spin-0 and spin-1 heavy diquark field is composed of two heavy quark fields
\begin{align}
&S^{i}(x)  =N_{S_{Q_1Q_2}} \epsilon^{ijk}Q_{1}{}_{j\beta}(x)Q_{2}{}_{k\gamma}(x)[C\gamma_{5}(1+\slashed
v)]_{\beta\gamma}~,\label{Scompose}\\
&X_{\mu}^{i}(x)  =N_{X_{Q_1Q_2}} \epsilon^{ijk}Q_{1}{}_{j\beta}(x)Q_{2}{}_{k\gamma}(x)[C\gamma_{\mu}(1+\slashed v)
]_{\beta\gamma}~.\label{Xcompose}
\end{align}
Thus the intepolating field of a doubly heavy baryon can be expressed by the combination of a diquark field
and a light quark field
\begin{align}
\Phi^{1/2(0)}_{Q_{1}Q_{2}}(x) & =\frac{N^{1/2(0)}_{Q_1Q_2}}{N_{S_{Q_1Q_2}}}\frac{1}{2}\bar{u}^{\alpha}q_{i\alpha}(x)S^{i}(x)~,
\label{Phi120}\\
\Phi^{1/2(1)}_{Q_{1}Q_{2}}(x) & =\frac{N^{1/2(1)}_{Q_1Q_2}}{N_{X_{Q_1Q_2}}}\frac{1}{2}[\bar{u}\gamma_{5}(\gamma^{\mu}
+v^{\mu})]^{\alpha}q_{i\alpha}(x)X_{\mu}^{i}(x)~,\\
\Phi^{3/2}_{Q_{1}Q_{2}}(x) & =\frac{N^{3/2}_{Q_1Q_2}}{N_{X_{Q_1Q_2}}}{\bar u}^{\mu, \alpha}q_{i\alpha}(x)X_{\mu}^{i}(x)~.
\label{Phi32}
\end{align}
In fact, these normalization factors are related by the heavy flavor symmetry, which leads to
\begin{align}
&N_{Q_{1}Q_{2}}^{1/2(1)}\to N^{1/2(1)},~~N_{Q_{1}Q_{2}}^{1/2(0)}\to N^{1/2(0)},~~N_{Q_{1}Q_{2}}^{3/2}\to N^{3/2}~,\nonumber\\
&N_{X_{Q_{1}Q_{2}}}\to N_X,~~N_{S_{Q_{1}Q_{2}}}\to N_S~.
\end{align}
However, the relation between $N_X$ and $N_S$  as well as the relation among $N^{1/2}, N^{1/2(0)}\  \text{and}\
N^{3/2}$ are not obvious. According to Eq.~(\ref{Scompose}) and Eq.~(\ref{Xcompose}), we can write the spinor
structure of the scalar and axial-vector diquarks in momentum space as
\begin{align}
S^{ss^{\prime}}(v) & =N_{S}u_{1\beta}^{s}(v)u_{2\gamma}^{s^{\prime}}(v)[C\gamma_{5}(1+\slashed v)]_{\beta\gamma}~,\nonumber\\
X_{\mu}^{ss^{\prime}}(v) & =N_{X}u_{1\beta}^{s}(v)u_{2\gamma}^{s^{\prime}}(v)[C\gamma_{\mu}(1+\slashed v)]_{\beta\gamma}~.
\end{align}
Here, we have omitted the color indices. $s, s^{\prime}$ denote the helicity of the spinors $u_1, u_2$, in order.
Since $X_{\mu}^{ss^{\prime}}$ has three independent degrees of freedom, while $S^{ss^{\prime}}$ has only
one degree of freedom, we can derive the following relation
\begin{equation}
\sum_{ss^{\prime}}S^{\dagger ss^{\prime}}(v)S^{ss^{\prime}}(v)=\frac{1}{3}\sum_{ss^{\prime}}g^{\mu\nu}
X_{\mu}^{\dagger ss^{\prime}}(v)X_{\nu}^{ss^{\prime}}(v)~,
\label{NXNSrela}
\end{equation}
where the sum of all the helicity indices is equivalent to counting the total degrees of freedom. The
relations among $N^{1/2}, N^{1/2(0)}\  \text{and}\  N^{3/2}$ can be determined by a similar approach.
We transform Eq.~(\ref{Phi120}-\ref{Phi32}) into the spinor structure in momentum space
\begin{align}
\Phi_{rlss^{\prime}}^{1/2(0)}(v) & =\frac{N^{1/2(0)}}{N_{S}}\frac{1}{2}\bar{u}^{r}(v)u^{l}(v)S^{ss^{\prime}}(v)~,\\
\Phi_{rlss^{\prime}}^{1/2(1)}(v) & =\frac{N^{1/2(1)}}{N_{X}}\frac{1}{2}\bar{u}^{r}(v)\gamma_{5}(\gamma^{\mu}+v^{\mu})
q^{l}(v)X_{\mu}^{ss^{\prime}}(v)~,\\
\Phi_{rlss^{\prime}}^{3/2}(v) & =\frac{N^{3/2}}{N_{X}}\bar{u}^{\mu,r}(v)u^{l}(v)X_{\mu}^{ss^{\prime}}(v)~,
\end{align}
where $r, l, s, s^{\prime}$ denote the helicities. Since a spin-1/2 particle has two degrees of freedom while a
spin-3/2 particle has four, we require the following relations
\begin{equation}
\sum_{rlss^{\prime}}\Phi_{rlss^{\prime}}^{1/2(0)\dagger}(v)\Phi_{rlss^{\prime}}^{1/2(0)}(v)=\sum_{rlss^{\prime}}
\Phi_{rlss^{\prime}}^{1/2\dagger}(v)\Phi_{rlss^{\prime}}^{1/2}(v)=\frac{1}{2}\sum_{rlss^{\prime}}\Phi_{rlss^{\prime}}^{3/2\dagger}(v)
\Phi_{rlss^{\prime}}^{3/2}(v)~.
\label{NX12NS12rela}
\end{equation} 
Finally, according to Eq.~(\ref{NXNSrela}) and Eq.~(\ref{NX12NS12rela}) we arrive at
\begin{equation}
N_S=\sqrt{2}N_X,~~~N^{1/2(0)}=\sqrt{6}N^{1/2(1)},~~~N^{3/2}=\frac{\sqrt{3}}{2}N^{1/2(1)}~,
\label{Nrelation}
\end{equation}
where the following properties have been used
\begin{equation}
\slashed v\ u=u,\ \ \slashed v\ u^{\mu}=u^{\mu},\ \ \ v_{\mu}u^{\mu}=0,\ \ \ \sum_{l}u^{l}\bar{u}^{l}=1+\slashed v,
\ \ \sum_{r}\bar{u}^{\mu,r}u_{\mu}^{r}=2\sum_{r}\bar{u}^{r}u^{r}~.
\end{equation}
It should be mentioned that the spinors used here are rescaled from the standard ones as ${\sqrt m_Q}u=u_{\rm QCD}$.
However, as long as we also choose rescaled states as ${\sqrt m_Q}|\cdots\rangle= |\cdots\rangle_{\rm QCD}$,
this will never affect our calculations.

\subsection{Transition Matrix Element}

With DiET, the transition matrix element defined in Eq.~(\ref{LSZformula}) can be calculated at the diquark level.
Further, in the heavy diquark limit, utilizing the technique given in Ref.~\cite{Hussain:1994zr},  we can
reduce the transition matrix element so that it will depend on less unknown form factors. 
Consider first the case of ${\cal B}_{bQ}^{1/2(1)}\to {\cal B}_{cQ}^{1/2(1)}$. The flavor changing current is
\begin{equation}
J^{b\to c}_{\mu}= X_{\rho,j}^{\dagger(c)}[\Gamma_{\mu}^{\rho\sigma}(\overleftarrow{\partial},\partial)]_{k}^{j}
X_{\sigma}^{(b)k}~,
\label{diqCurr}
\end{equation}
where $j,\ k$ are color indices. $[\Gamma_{\mu}^{\rho\sigma}]_{k}^{j}$ can be factorized as $\Gamma_{\mu}^{\rho\sigma}
\times C \delta_{k}^{j}$, and $C=-1/3$ is given in the last section from matching. To leading power of $1/m_{X}^{2}$,
one can approximate the $X_{\mu}$ field as ${X}_{v\mu}$, so that the $\overleftarrow{\partial},\ \partial$ in
Eq.~(\ref{diqCurr})
can be replaced with $i v_2,\ -i v_1$. According to the reduction formula Eq.~(\ref{LSZformula}),
the transition matrix element in DiET is
\begin{align}
& \langle{\cal B}_{cQ}^{1/2(1)}(P_c)|J^{b\to c}_{\mu}(0)|{\cal B}_{bQ}^{1/2(1)}(P_b)\rangle= \frac{(N^{1/2(1)})^2}{N_{X}^2}L(P_{b},P_{c})\int d^{4}xd^{4}y\ e^{i(P_{c}-m_{X_c}v_{2})\cdot x}e^{-i(P_{b}-m_{X_b}v_{1})\cdot y}\nonumber\\
  & \times \bar{\chi}_{\alpha,a}^{(c)}\chi_{\beta,b}^{(b)}C\langle0|T\big\{X_{v_{2}(c)}^{\alpha,i}(x)q_{i}^{a}(x)\ X_{v_2\rho,j}^{\dagger(c)}(0)\Gamma_{\mu}^{\rho\sigma}X_{v_1\sigma}^{(b)j}(0)\ X_{v_1(b)l}^{\dagger\beta}(y)\bar{q}^{l,b}(y)\big\}|0\rangle~,
\label{MatrixLSZ}
\end{align}
where $a,\ b$ are Dirac indices, while $i,\ j,\ k,\ l$ are color indices. $v_1$ and $v_2$ are is four-velocity of
the initial and the final baryon, respectively. Using the decoupling transformation defined in
Eq.~(\ref{decoTrans}), and noting that the $\tilde X_{v\mu}$ fields are totally decoupled from the soft
gluons and also the light quarks, one can factorize the time-ordered matrix element in Eq.~(\ref{MatrixLSZ}) to be
\begin{align}
&\langle0|T\Big\{W\Big[\begin{array}{c}
x\\
v_{2}
\end{array}\Big]_{i^{\prime}}^{i}W^{-1}\Big[\begin{array}{c}
0\\
v_{2}
\end{array}\Big]_{j}^{j^{\prime}}W\Big[\begin{array}{c}
0\\
v_{1}
\end{array}\Big]_{k^{\prime}}^{j}W^{-1}\Big[\begin{array}{c}
y\\
v_{1}
\end{array}\Big]_{l}^{l^{\prime}}q_{i}^{a}(x)\bar{q}^{l,b}(y)\Big\}|0\rangle\nonumber\\
 & \times\langle0|T\big\{\tilde{X}_{v_2(c)}^{\alpha,i^{\prime}}(x)\tilde{X}_{v_2\rho,j^{\prime}}^{\dagger(c)}(0)\big\}|0\rangle\Gamma_{\mu}^{\rho\sigma}\langle0|T\big\{\tilde{X}_{v_1\sigma(b)}^{k^{\prime}}(0)\tilde{X}_{v_1l^{\prime}}^{\dagger\beta(b)}(y)\big\}|0\rangle~.\label{SoftFunc1}
\end{align}
The last two matrix elements in Eq.~(\ref{SoftFunc1}) can be calculated directly from the free diquark
propagator Eq.~(\ref{FreeDiProp}).
Using the fact that $\bar{\chi}_{\alpha,a}^{(c)}v_2^{\alpha}=\chi_{\beta,b}^{(b)}v_1^{\beta}=0$, one has
\begin{align}
 &\langle{\cal B}_{cQ}(P_c)^{1/2(1)}|J^{b\to c}_{\mu}(0)|{\cal B}_{bQ}^{1/2(1)}(P_b)\rangle\nonumber\\
 = & -C\frac{(N^{1/2(1)})^2}{N_{X}^2}\frac{1}{4m_{X_c}m_{X_b}}L(P_{b},P_{c})\int d^{4}kd^{4}q[\bar{u}\gamma_{5}(\gamma_{\alpha}+v_{2\alpha})]_{a}\Gamma_{\mu}^{\alpha\beta}[(\gamma_{\beta}+v_{1\beta})\gamma_{5}u]_{b}\nonumber\\
 & \times M(k,q;v_{2},v_{1})^{ab}\frac{1}{v_{2}\cdot(P_{c}-m_{X}v_{2}-k)}\frac{1}{v_{1}\cdot(P_{b}-m_{X}v_{1}+q)}~.
 \label{MatrixEle1}
\end{align}
The dynamics of the light degrees of freedom is completely encapsulated in the following Fourier transformed
soft function
\begin{align}
&M(k,q;v_{2},v_{1})^{ab} =\int\frac{d^{4}k}{(2\pi)^{4}}\frac{d^{4}q}{(2\pi)^{4}}e^{-ik\cdot x}e^{-iq\cdot y}\nonumber\\
&\times\langle0|T\Big\{W\Big[\begin{array}{c}
x\\
v_{2}
\end{array}\Big]_{i^{\prime}}^{i}W^{-1}\Big[\begin{array}{c}
0\\
v_{2}
\end{array}\Big]_{j}^{i^{\prime}}W\Big[\begin{array}{c}
0\\
v_{1}
\end{array}\Big]_{k^{\prime}}^{j}W^{-1}\Big[\begin{array}{c}
y\\
v_{1}
\end{array}\Big]_{l}^{k^{\prime}}q_{i}^{a}(x)\bar{q}^{l,b}(y)\Big\}|0\rangle~.
\label{softfunc}
\end{align}

Next, we need to extract the residues of the mass poles by applying the operator $L(P_{b},P_{c})$ on
the correlation function.
Near the mass shell, the external momenta $P_{Q}$ can be parameterized as
\begin{align}
& P_{Q}  =M_{Q}(1+\epsilon_{Q})v_{Q}+M_{Q}\epsilon_{\perp},\ \ \ (\epsilon_{\perp}\cdot v_Q=0)\nonumber\\
& L(P_{Q})  =\lim_{\epsilon\to0}(P_{Q}^{2}-M_{Q}^{2})=\lim_{\epsilon\to0}(2\epsilon_{Q}+\epsilon_{\perp}^2)M_{Q}^{2}~.
\end{align}
Although the decoupling transformation Eq.~(\ref{decoTrans}) realizes the factorization as shown
in Eq.~(\ref{SoftFunc1}), there still exist non-perturbative interactions between the heavy and light
degrees of freedom due to  confinement. Such
effects have been absorbed into the momentum distribution of $M(k,q;v_{2},v_{1})$. In other words,
the light particles in the
baryon always ``know'' that they are bound with a heavy diquark. To reflect the confinement,
$M(k,q;v_{2},v_{1})$ is assumed to  
peak at $v_{2}\cdot k=\bar{\Lambda}_{c},\  v_{1}\cdot q=-\bar{\Lambda}_{b}$, where $\bar{\Lambda}_{Q}=M_{Q}-m_{X_{Q}}$.
Operating with $L(P_{Q})$ on the denominators, taking the limit $\epsilon_Q, \epsilon_{\perp} \to 0$, and
noting that there are no poles of $1/\epsilon_{\perp}^2$, one gets
\begin{align}
L(P_{c})\frac{1}{v_{2}\cdot(P_{c}-m_{X_c}v_2-k)}  = 2M_c~,~~~
L(P_{b})\frac{1}{v_{1}\cdot(P_{b}-m_{X_b}v_1+q)}  =2M_b~.
\end{align}
On the other hand, the soft function can be generally parametrized as
\begin{equation}
\int d^{4}kd^{4}qM(k,q;v_{2},v_{1})^{ab}=[A(w)+B(w)\slashed v_{1}+C(w)\slashed v_{2}+D(w)\slashed v_{2}
\slashed v_{1}]^{ab}~.
\end{equation}
However, the $B(w),\ C(w),\ D(w)$ form factors can be totally absorbed into the the form factor $A(w)$
since $\bar{u}_{c}\gamma_{5}(\gamma_{\alpha}+v_{2\alpha})\slashed v_{2}=\bar{u}_{c}\gamma_{5}(\gamma_{\alpha}
+v_{2\alpha})$ and $\slashed v_{1}(\gamma_{\beta}+v_{1\beta})\gamma_{5}u=(\gamma_{\beta}+v_{1\beta})\gamma_{5}u$,
which leaves only one $w$-dependent form factor denoted as $A^{\prime}(w)$. Explicitly they are related
by $A^{\prime}(w)={\cal F}[A(w),B(w),C(w),D(w)]$. Thus we have
\begin{align}
\text{} & \langle{\cal B}_{cQ}^{1/2(1)}(P_c)|J^{b\to c}_{\mu}(0)|{\cal B}_{bQ}^{1/2(1)}(P_b)\rangle\nonumber\\
= & -C\frac{(N^{1/2(1)})^2 M^2}{N_{X}^2m_X^2}A^{\prime}(w)[\bar{u}\gamma_{5}(\gamma_{\alpha}+v_{2\alpha})]\Gamma_{\mu}^{\alpha\beta}[(\gamma_{\beta}+v_{1\beta})\gamma_{5}u]~,
\label{12Ga12}
\end{align}
where the masses are blind to the flavors so that $M_b=M_c=M$ and $m_{X_{b}}=m_{X_{c}}=m_X$. Similarly, for
the 1/2(1) $\to$ 1/2(0), 1/2(0) $\to$ 1/2(1) and 1/2(1) $\to$ 3/2(1) transitions, we have
\begin{align}
&\langle{\cal B}^{1/2(0)}_{cQ}(P_c)|J^{b\to c}_{\mu}(0)|{\cal B}^{1/2(1)}_{bQ}(P_b)\rangle= \sqrt{3}C\frac{(N^{1/2(1)})^2 M^2}{N_{X}^2m_X^2}A^{\prime}(w)\ \bar{u}\Gamma_{\mu}^{\beta}(\gamma_{\beta}+v_{1\beta})\gamma_{5}u~,\label{12SGa12}\\
&\langle{\cal B}^{1/2(1)}_{cQ}(P_c)|J^{b\to c}_{\mu}(0)|{\cal B}^{1/2(0)}_{bQ}(P_b)\rangle= \sqrt{3}C\frac{(N^{1/2(1)})^2 M^2}{N_{X}^2m_X^2}A^{\prime}(w)\ \bar{u}\gamma_{5}(\gamma_{\alpha}+v_{2\alpha})\Gamma_{\mu}^{\alpha}u~,\label{12SGa12}\\
&\langle{\cal B}^{3/2(1)}_{cQ}(P_c)|J^{b\to c}_{\mu}(0)|{\cal B}^{1/2(1)}_{bQ}(P_b)\rangle= -\sqrt{3}C\frac{(N^{1/2(1)})^2 M^2}{N_{X}^2 m_X^2}A^{\prime}(w)\ \bar{u}_{\alpha}\Gamma_{\mu}^{\alpha\beta}(\gamma_{\beta}+v_{1\beta})\gamma_{5}u~,\label{32Ga12}
\end{align}
where Eq.~(\ref{Nrelation}) has been used. The unknown function $A^{\prime}(w)$ contains all the dynamics
of light degrees of freedom, and it describes the response of the light particles to the changing of
heavy diquark velocity.
Furthermore,  $A^{\prime}(w)$ is totally determined by the soft function Eq.~(\ref{softfunc}). In fact,
this soft function is
a universal quantity which also appears in the HQET analysis of $B\to D$ transition \cite{Hussain:1994zr}, where
the Isgur-Wise function $\xi(w)$ is derived from it in the same way as done here for $A^{\prime}(w)$.
Explicitly, ${\xi}(w)\propto{\cal F}[A(w),B(w),C(w),D(w)]$. Thus one can conclude that $A^{\prime}(w)$
is related  to $\xi(w)$   up to some constant coefficients.

\subsection{Phenomenological Results for Reduced Form Factors}
Generally, the doubly heavy baryon transition matrix element induced by the $V-A$ current is parametrized by
several independent form factors. For ${\cal B}_{bQ}^{1/2}\to {\cal B}_{cQ}^{1/2}$ it reads
\begin{eqnarray}
&&\langle{\cal B}_{cQ}^{1/2}(P_c)|\left(J_{\mu}^{V}(0)-J_{\mu}^{A}(0)\right)|{\cal B}_{bQ}^{1/2}(P_b)\rangle
\nonumber \\ & = & \bar{u}_{cQ}(P_{c})\bigg[F_{1}(q^{2})\gamma^{\mu}+F_{2}(q^{2})P_{c}^{\mu}+F_{3}(q^{2})
P_{b}^{\mu}\bigg]u_{bQ}(P_{b})\nonumber \\
&  &- \bar{u}_{cQ}(P_{c})\bigg[G_{1}(q^{2})\gamma^{\mu}+G_{2}(q^{2})P_{c}^{\mu}+G_{3}(q^{2})P_{b}^{\mu}\bigg]
\gamma_{5}u_{bQ}(P_{b})~, 
\label{eq:parameterization1}
\end{eqnarray}
while for ${\cal B}_{bQ}^{1/2}\to {\cal B}_{cQ}^{3/2}$ the parametrization takes the form
\begin{align}
& \langle{\cal B}_{cQ}^{3/2}(P_{c})|\left(J_{\mu}^{V}(0)-J_{\mu}^{A}(0)\right)|{\cal B}_{bQ}^{1/2}(P_{b})\rangle
\nonumber\nonumber \\
= \ &\bar{u}_{cQ}^{\alpha}(P_{c})\bigg[\frac{f_{1}^{\prime}(q^{2})}{M_{b}}\gamma^{\mu}P_{b\alpha}
+\frac{f_{2}^{\prime}(q^{2})}{M_{b}^{2}}P_{b\alpha}P_{b\mu}+\frac{f_{3}^{\prime}(q^{2})}{M_{b}M_{c}}P_{b\alpha}P_{c\mu}
+f_{4}^{\prime}(q^{2})g_{\mu\alpha}\bigg]\gamma_{5}u_{bQ}(P_{b})\nonumber\nonumber \\
& -\bar{u}_{cQ}^{\alpha}(P_{c})\bigg[\frac{g_{1}^{\prime}(q^{2})}{M_{b}}\gamma^{\mu}P_{b\alpha}
+\frac{g_{2}^{\prime}(q^{2})}{M_{b}^{2}}P_{b\alpha}P_{b\mu}+\frac{g_{3}^{\prime}(q^{2})}{M_{b}M_{c}}P_{b\alpha}
P_{c\mu}+g_{4}^{\prime}(q^{2})g_{\mu\alpha}\bigg]u_{bQ}(P_{b})~.
\end{align}
However, if we treat such process by HDiET  considering also the heavy flavor symmetry, the number of
independent form factors can be greatly reduced. Especially, by combining Eqs.~(\ref{XXCurrSym}-\ref{XSCurrSym})
and Eqs.~(\ref{12Ga12}-\ref{32Ga12}), one arrives at
\begin{align}
&\langle{\cal B}^{1/2(1)}_{cQ}|J^{V}_{\mu}(0)|{\cal B}^{1/2(1)}_{bQ}\rangle= \eta(w)\bar{u}\big[2(1+w)\gamma_{\mu}+v_{b\mu}+v_{c\mu}\big]u~,\label{FF12V12}\\
&\langle{\cal B}^{1/2(1)}_{cQ}|J^{A}_{\mu}(0)|{\cal B}^{1/2(1)}_{bQ}\rangle= \eta(w)\bar{u}\big[2(1+w)\gamma_{\mu}\big]\gamma_{5}u~,\label{FF12A12}\\
&\langle{\cal B}^{1/2(0)}_{cQ}|J^{V}_{\mu}(0)|{\cal B}^{1/2(1)}_{bQ}\rangle= -\sqrt{3}\eta(w)\bar{u}\big[(1+w)\gamma_{\mu}-v_{b\mu}-v_{c\mu}\big]u~,\label{FF120V12}\\
&\langle{\cal B}^{1/2(0)}_{cQ}|J^{A}_{\mu}(0)|{\cal B}^{1/2(1)}_{bQ}\rangle= \sqrt{3}\eta(w)\bar{u}\big[(1+w)\gamma_{\mu}\big]\gamma_5 u~,\label{FF120A12}\\
&\langle{\cal B}^{1/2(1)}_{cQ}|J^{V}_{\mu}(0)|{\cal B}^{1/2(0)}_{bQ}\rangle= \sqrt{3}\eta(w)\bar{u}\big[(1+w)\gamma_{\mu}-v_{b\mu}-v_{c\mu}\big]u~,\label{FF12V120}\\
&\langle{\cal B}^{1/2(1)}_{cQ}|J^{A}_{\mu}(0)|{\cal B}^{1/2(0)}_{bQ}\rangle= \sqrt{3}\eta(w)\bar{u}\big[(1+w)\gamma_{\mu}\big]\gamma_5 u~,\label{FF12A120}\\
&\langle{\cal B}^{3/2(1)}_{cQ}|J_{\mu}^{V}(0)|{\cal B}^{1/2(1)}_{bQ}\rangle =-\sqrt{3}\eta(w)\bar{u}_{\alpha}\big[(1+w)g^{\alpha}_{\mu}-v_{c\mu}v_{b}^{\alpha}+\gamma_{\mu}v_{b}^{\alpha}\big]\gamma_{5}u~,\label{FFV32}\\
&\langle{\cal B}^{3/2(1)}_{cQ}|J_{\mu}^{A}(0)|{\cal B}^{1/2(1)}_{bQ}\rangle =\sqrt{3}\eta(w)\bar{u}_{\alpha}\big[(1+w)g^{\alpha}_{\mu}-v_{c\mu}v_{b}^{\alpha}\big]u~, \label{FFA32}
\end{align}
where only one form factor $\eta(w)$ is left. This  is shared by all the six matrix elements and $\eta(w)$ is
proportional to the soft function $A^{\prime}(w)$
\begin{equation}
\eta(w)=C\Lambda\frac{(N^{1/2})^2 M^2}{N_{X}^2m_X^2}A^{\prime}(w)~.
\label{etaFunc}
\end{equation}
The vector transition shown in Eq.~(\ref{FF12V12}) is exactly the same as that given in~\cite{Carone:1990pv},
where the transition
matrix element was derived based on heavy quark-diquark symmetry. However, Ref.~\cite{Carone:1990pv} did not
give the result for the axial-current transition. In terms of the complicated factors in Eq.~(\ref{etaFunc}),
this is determined through the  normalization at the zero-recoil point $w=1$. From Eq.~(\ref{XXCurrSym}),
one can find that
the vector current $J_{\mu(V)}^{X\to X}$ is conserved $\partial^{\mu}J_{\mu(V)}^{X\to X}=0$. This implies
the conservation of diquark number. Thus we can conclude that
\begin{align}
\langle{\cal B}_{cQ}(v)|\int d^3\vec x\ J_{0(V)}^{X\to X}(\vec x)|{\cal B}_{bQ}(v)\rangle
= \langle{\cal B}_{cQ}(v)|\mathbf{1}|{\cal B}_{bQ}(v) \rangle=2v^0(2\pi)^3\delta^3(0)~,
\end{align}
where $\mathbf{1}$ means the diquark number is one. On the other hand, using Eq.~(\ref{FF12V12}), and choosing
the rest-frame of ${\cal B}_{bQ}(v)$, $v=(1,\vec 0)$, the same matrix element becomes
\begin{align}
\langle{\cal B}_{cQ}(v)|\int d^3\vec x\ J_{0(V)}^{X\to X}(\vec x)|{\cal B}_{bQ}(v)\rangle
&=(2\pi)^3\delta^3(0)\eta(1)\bar{u}(v)\big[4\gamma_{0}+2v_{0}\big]u(v)\nonumber\\
&=12\eta(1)v^0(2\pi)^3\delta^3(0)~,
\end{align}
where we have used $\gamma_0=\slashed v$ and $\slashed v u=u$. Comparing the above two equations,
one can conclude that $\eta(1)=1/6$. At the end of last subsection, we have argued that $A^{\prime}(w)
\propto\xi(w)$. Since $\xi(1)=1$, it thus follows that $\eta(1)=(1/6)\xi(1)$. 

However, it is necessary to point out  that the reduced matrix elements Eqs.~(\ref{FF12V12}-\ref{FFA32}) are
only applicable in the region $w\sim 1$ or equivalently $q^2\sim q_{max}^2=(M_b-M_c)^2$. In the smaller-$q^2$
region, the large recoil may invalidate the static dynamics of HDiET. As a result, one cannot argue that
for any $w$ we have $\eta(w)=(1/6)\xi(w)$, and an appropriate extension of the form factors
from $q^2=q_{max}^2$ to $q^2=0$ is necessary.
Since the transition matrix elements Eqs.~(\ref{FF12V12}-\ref{FFA32}) are expected to have a lowest-$q^2$ pole
at the mass of $B_c$ meson, it is appropriate to multiply $\eta(w)$ with single pole function $B(w)$ with a
suitable normalization $B(1)=1$, 
 \begin{align}
B\left(w=\frac{M_b^2+M_c^2-q^2}{2M_bM_c}\right)=\frac{1-q_{max}^2/m_{B_c}^2}{1-q^2/m_{B_c}^2}~.
\end{align}
Finally, we arrive at an explicit expression of the $\eta$ function
\begin{align}
\eta(q^2)=\frac{1}{6}\xi\left(\frac{M_b^2+M_c^2-q^2}{2M_bM_c}\right)\frac{1-q_{max}^2/m_{B_c}^2}{1-q^2/m_{B_c}^2}~.
\label{etafunc}
\end{align}
Note that for the practical calculation we have to distinguish bewteen the different masses $M_b, M_c$.
The Isgur-Wise function was calculated e.g. in Ref.~\cite{Faller:2008tr}, which has the expression
\begin{align}
\xi(w)=\int\limits_0^{\beta_0/w} &d\rho 
\exp\left(\frac{\bar{\Lambda}-\rho w}{\tau}\right)
\Big[\frac{1}{2w}\phi_{-}^B(\rho)+
\big(1-\frac{1}{2w}\big)\phi_{+}^B(\rho) \Big]~,\\
s_0^D&=\kappa^2 m_Q^2+2\kappa m_Q\beta_0,~~
M^2=2\kappa m_Q\tau~,\label{scal} 
\end{align}
where $\bar \Lambda=m_B-m_b$, $m_Q=m_b$,  $\kappa=m_c/m_b$,  $s_0^D=6\ \rm{GeV}^2$ is the effective
threshold, while $M^2=3-6$~GeV$^2$ is
the Borel parameter. In this work, we simply use its center value $M^2=4.5$ GeV$^2$. $\phi_{\pm}^B$
are the $B$ meson light-cone distribution amplitudes, which have the form
\begin{align}
\phi_+^B(\omega) = \dfrac{\omega}{\omega_0^2}\,e^{-\frac{\omega}{\omega_0}},\ \ \ 
\phi_-^B(\omega) = \dfrac{1}{\omega_0}\,e^{-\frac{\omega}{\omega_0}},
\label{eq-GN}
\end{align}
where $\omega_0=(2/3)\bar \Lambda$ \cite{Grozin:1996pq}. The  mass parameters are set as $m_b=4.18$~GeV,
$m_c=1.27$~GeV, $m_B=5.279$~GeV, $m_D=1.869$~GeV and $m_{Bc}=6.275$~GeV.
\begin{figure}
\includegraphics[width=1.0\columnwidth]{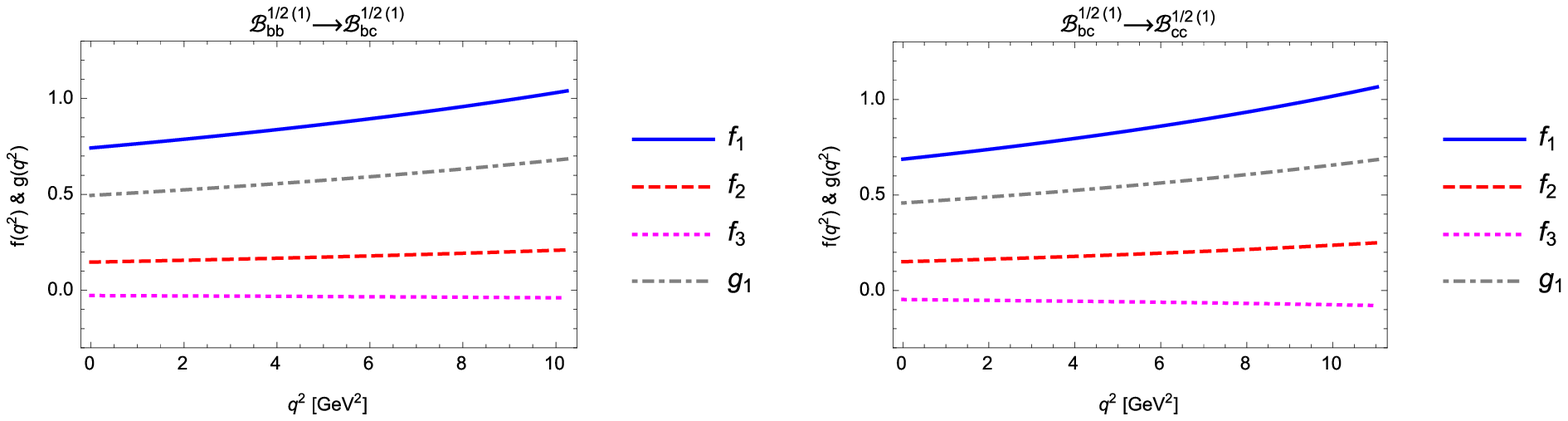} 
\includegraphics[width=1.0\columnwidth]{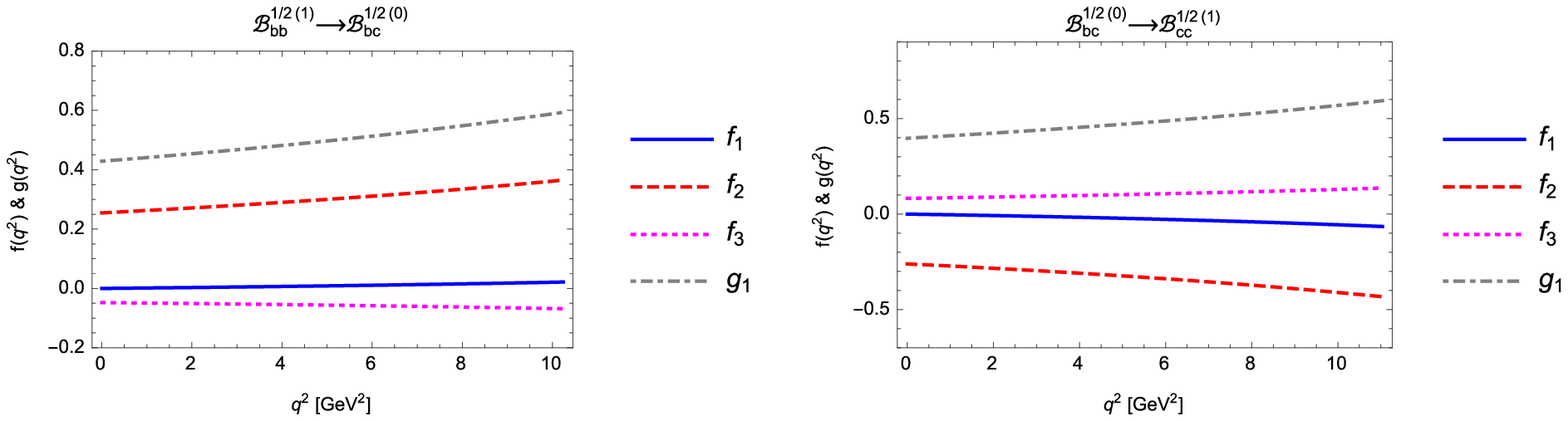} 
\caption{$q^2$-dependence of the ${\cal B}_{bQ}^{1/2}\to {\cal B}_{cQ}^{1/2}$ form factors, where $Q=b,c$.}
\label{fig:FFbQ12cQ12} 
\end{figure}
\begin{figure}
\includegraphics[width=1.0\columnwidth]{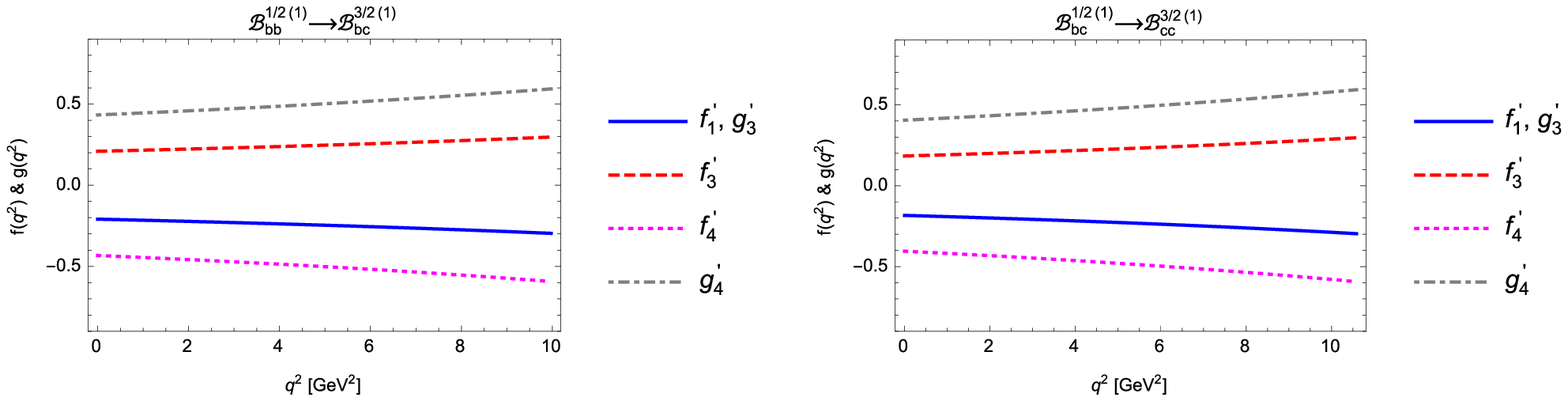} 
\caption{$q^2$-dependence of the ${\cal B}_{bQ}^{1/2(1)}\to {\cal B}_{cQ}^{3/2(1)}$ form factors, where $Q=b,c$.}
\label{fig:FFbQ12cQ32} 
\end{figure}
Fig.~\ref{fig:FFbQ12cQ12} shows $q^2$-dependence of the ${\cal B}_{bQ}^{1/2(1)}\to {\cal B}_{cQ}^{1/2(1)},
{\cal B}_{cQ}^{1/2(0)}$  form factors, where we have redefined the six form factors as
\begin{eqnarray}
&&\langle{\cal B}_{cQ}(P_c)|\left(J_{\mu}^{V}(0)-J_{\mu}^{A}(0)\right)|{\cal B}_{bQ}(P_b)\rangle \nonumber \\ & =
& \bar{u}_{cQ}(P_{c})\bigg[\gamma^{\mu}f_{1}(q^{2})+i\sigma^{\mu\nu}\frac{q_{\nu}}{M_b}f_{2}(q^{2})+\frac{q^{\mu}}{M_b}
f_{3}(q^{2})\bigg]u_{bQ}(P_{b})\nonumber \\
&  &- \bar{u}_{cQ}(P_{c})\bigg[\gamma^{\mu}g_{1}(q^{2})+i\sigma^{\mu\nu}\frac{q_{\nu}}{M_b}g_{2}(q^{2})+\frac{q^{\mu}}{M_b}g_{3}(q^{2})\bigg]
\gamma_{5}u_{b}(P_{b})~,
\label{eq:parameterization2}
\end{eqnarray}
with $q^{\mu}=P_{b}^{\mu}-P_{c}^{\mu}$ the transferred momentum. The $f_i$ and $g_i$ are related to the $F_i$
and $G_i$ as
\begin{eqnarray}
f_1(q^2) &=& F_1(q^2)+\frac{1}{2}(M_b+M_c)(F_2(q^2)+F_3(q^2))~, \nonumber \\
f_2(q^2) &=& \frac{1}{2}M_b(F_2(q^2)+F_3(q^2)),~~~f_3(q^2) = \frac{1}{2}M_b(F_3(q^2)-F_2(q^2))~, \nonumber \\
g_1(q^2) &=& G_1(q^2)-\frac{1}{2}(M_b-M_c)(G_2(q^2)+G_3(q^2))~, \nonumber \\
g_2(q^2) &=& \frac{1}{2}M_b(G_2(q^2)+G_3(q^2)),~~~g_3(q^2) = \frac{1}{2}M_b(G_3(q^2)-G_2(q^2))~,
\end{eqnarray}
and here we have $g_2(q^2)=g_3(q^2)=0$. The masses of the baryons are $m_{{\cal B}_{bb}}=10.143$~GeV,
$m_{{\cal B}_{bc}^{1/2}}=6.943$~GeV, $m_{{\cal B}_{bc}^{3/2}}=6.985$~GeV, $m_{{\cal B}_{cc}^{1/2}}=3.621$~GeV and
$m_{{\cal B}_{cc}^{3/2}}=3.69$~GeV. Fig.~\ref{fig:FFbQ12cQ32} shows the $q^2$-dependence of the
${\cal B}_{bQ}^{1/2(1)}\to {\cal B}_{cQ}^{3/2(1)}$  form factors, where $f_2^{\prime}(q^2)=g_1^{\prime}(q^2)
=g_2^{\prime}(q^2)=0$.

\subsection{Semi-Leptonic Decay Widths}

Next, using the form factors given in the last section, we will calculate the semi-leptonic decay
widths of ${\cal B}_{bQ}^{1/2(1)}\to {\cal B}_{cQ}^{1/2(1)}, {\cal B}_{cQ}^{1/2(0)}~\text{and}~{\cal B}_{cQ}^{3/2(1)}$.
For the case of ${\cal B}_{bQ}^{1/2(1)}\to {\cal B}_{cQ}^{1/2(1)}, {\cal B}_{cQ}^{1/2(0)}$, the formula of
the differential decay width is given in \cite{Wang:2017mqp,Shi:2019fph}
\begin{align}
\frac{d\Gamma_{L}}{dq^{2}} =&\ \frac{G_{F}^{2}|V_{{\rm CKM}}|^{2}q^{2}\ p\ (1-\hat{m}_{l}^{2})^{2}}{384\pi^{3}
M_{1}^{2}}\nonumber\\
&\ \times\left((2+\hat{m}_{l}^{2})(|H_{-\frac{1}{2},0}|^{2}+|H_{\frac{1}{2},0}|^{2})+3\hat{m}_{l}^{2}(|H_{-\frac{1}{2},t}|^{2}
+|H_{\frac{1}{2},t}|^{2})\right)~,
\label{eq:longi-1}\\
\frac{d\Gamma_{T}}{dq^{2}} =&\ \frac{G_{F}^{2}|V_{{\rm CKM}}|^{2}q^{2}\ p\ (1-\hat{m}_{l}^{2})^{2}(2+\hat{m}_{l}^{2})}{384\pi^{3}M_{1}^{2}}
(|H_{\frac{1}{2},1}|^{2}+|H_{-\frac{1}{2},-1}|^{2})~,
\label{eq:trans-1}
\end{align}
with the helicity amplitudes given as $H_{\lambda_{2},\lambda_{W}}=H_{\lambda_{2},\lambda_{W}}^{V}-H_{\lambda_{2},\lambda_{W}}^{A}$,
\begin{align}
& H_{\frac{1}{2},0}^{V}  = -i\frac{\sqrt{Q_{-}}}{\sqrt{q^{2}}}\left((M_{1}+M_{2})f_{1}-\frac{q^{2}}{M_{1}}f_{2}\right)~,\;\;\;
H_{\frac{1}{2},0}^{A} =  -i\frac{\sqrt{Q_{+}}}{\sqrt{q^{2}}}\left((M_{1}-M_{2})g_{1}+\frac{q^{2}}{M_1}g_{2}\right)~,\nonumber \\
& H_{\frac{1}{2},1}^{V}  = i\sqrt{2Q_{-}}\left(-f_{1}+\frac{M_{1}+M_{2}}{M_{1}}f_{2}\right)~,\;\;\;
H_{\frac{1}{2},1}^{A}  =  i\sqrt{2Q_{+}}\left(-g_{1}-\frac{M_{1}-M_{2}}{M_{1}}g_{2}\right)~,\nonumber \\
& H_{\frac{1}{2},t}^{V}  = -i\frac{\sqrt{Q_{+}}}{\sqrt{q^{2}}}\left((M_{1}-M_{2})f_{1}+\frac{q^{2}}{M_{1}}f_{3}\right)~,\;\;\;
H_{\frac{1}{2},t}^{A} =  -i\frac{\sqrt{Q_{-}}}{\sqrt{q^{2}}}\left((M_{1}+M_{2})g_{1}-\frac{q^{2}}{M_{1}}g_{3}\right)~, \nonumber \\
&H_{-\lambda_{2},-\lambda_{W}}^{V}=H_{\lambda_{2},\lambda_{W}}^{V}~,\quad H_{-\lambda_{2},-\lambda_{W}}^{A}=-H_{\lambda_{2},\lambda_{W}}^{A}~,\quad\text{and}\quad Q_{\pm}=(M_1\pm M_2)^{2}-q^{2}~,
\end{align}
where $p=M_2\sqrt{w^2-1}$ with $w=(M_1^2+M_2^2-q^2)/2M_1M_2$, while $M_1(M_2)$ is the initial(final) baryon mass,
$\hat{m}_{l}\equiv m_{l}/\sqrt{q^{2}}$ and $m_l$ is the lepton mass. For the case of ${\cal B}_{bQ}^{1/2(1)}
\to {\cal B}_{cQ}^{3/2(1)}$, the formula for the differential decay width is given in~\cite{Zhao:2018mrg}
\begin{align}
\frac{d\Gamma_{T}}{dw} & =  \frac{G_{F}^{2}}{(2\pi)^{3}}|V_{{\rm CKM}}|^{2}\frac{q^{2}M_1^{\prime2}\sqrt{w^{2}-1}}{12M}\left(|{\bar H}_{{1\over 2},1}|^{2}+|{\bar H}_{-{1\over 2},-1}|^{2}+|{\bar H}_{{3\over 2},1}|^{2}+|{\bar H}_{-{3\over 2},-1}|^{2}\right)~,\\
\frac{d\Gamma_{L}}{dw} & =  \frac{G_{F}^{2}}{(2\pi)^{3}}|V_{{\rm CKM}}|^{2}\frac{q^{2}M_1^{\prime2}\sqrt{w^{2}-1}}{12M}\left(|{\bar H}_{{1\over 2},0}|^{2}+|{\bar H}_{-{1\over 2},0}|^{2}\right)~,
\end{align}
with the helicity amplitudes given as
\begin{eqnarray}
{\bar H}_{3/2,1}^{V,A} & = & \mp i\sqrt{2M_1M_2(w\mp1)}f_{4}^{\prime V,A}~,\\
{\bar H}_{1/2,1}^{V,A} & = & i\sqrt{\frac{2}{3}}\sqrt{M_1M_2(w\mp1)}\left[f_{4}^{\prime V,A}-2(w\pm1)f_{1}^{\prime V,A}\right]~,\\
{\bar H}_{1/2,0}^{V,A} & = & \pm i\frac{1}{\sqrt{q^{2}}}\frac{2}{\sqrt{3}}\sqrt{M_1M_2(w\mp1)}\Big[(M_1w-M_2)f_{4}^{\prime V,A}\mp(M_1\mp M_2)(w\pm1)f_{1}^{\prime V,A}\nonumber \\
&  & \qquad\qquad\qquad\qquad\qquad\qquad+M^{\prime}(w^{2}-1)f_{2}^{\prime V,A}+M(w^{2}-1)f_{3}^{\prime V,A}\Big]~,
\end{eqnarray}
where the upper (lower) sign denotes $V$ ($A$), $f_{i}^{V}=f_{i}$
($f_{i}^{A}=g_{i}$). The total decay
width is the sum of the longitudinal and the transversal parts
\begin{equation}
\Gamma=\int_{m_l^2}^{(M_{1}-M_{2})^{2}}dq^{2}\left(\frac{d\Gamma_{L}}{dq^{2}}+\frac{d\Gamma_{T}}{dq^{2}}\right)~.
\end{equation}
 The masses of $e, \mu$ are
 neglected here and $m_{\tau}=1.78$~GeV. Tab.~\ref{Tab:semi_width} gives the resulting decay widths and also
 a comparison with those derived in Ref.\cite{Wang:2017mqp} within light-front quark model (LFQM).
 It appears that the two sets of decay width results are consistent.
\begin{table}
\caption{Decay widths of ${{\cal B}}_{bb}\to{{\cal B}}_{bc}(l/\tau)\nu$. Comparison
between the results in this work and those derived in Ref.~\cite{Wang:2017mqp}
using the LFQM. }
\label{Tab:semi_width} %
\begin{tabular}{c|c|c|c|c}
\hline 
\hline 
Channel  & $\Gamma[{\rm GeV}]$ (This Work)  & $\Gamma_{L}/\Gamma_{T}$ (This Work)  & $\Gamma[{\rm GeV}]$ (LFQM)  & $\Gamma_{L}/\Gamma_{T}$ (LFQM)\tabularnewline
\hline 
${\cal B}_{bb}^{1/2(1)}\to{\cal B}_{bc}^{1/2(1)}l\nu$  & $4.1\times10^{-14}$  & $2.54$  & $3.3\times10^{-14}$  & $2.32$\tabularnewline
${\cal B}_{bb}^{1/2(1)}\to{\cal B}_{bc}^{1/2(1)}\tau\nu$  & $1.0\times10^{-14}$  & $2.12$  &  & \tabularnewline
${\cal B}_{bc}^{1/2(1)}\to{\cal B}_{cc}^{1/2(1)}l\nu$  & $3.2\times10^{-14}$  & $2.47$  & $4.5\times10^{-14}$  & $2.48$\tabularnewline
${\cal B}_{bc}^{1/2(1)}\to{\cal B}_{cc}^{1/2(1)}\tau\nu$  & $0.9\times10^{-14}$  & $2.13$  &  & \tabularnewline
${\cal B}_{bb}^{1/2(1)}\to{\cal B}_{bc}^{1/2(0)}l\nu$  & $1.8\times10^{-14}$  & $1.12$ & $1.5\times10^{-14}$ & 0.91\tabularnewline
${\cal B}_{bb}^{1/2(1)}\to{\cal B}_{bc}^{1/2(0)}\tau\nu$  & $4.4\times10^{-15}$ & $0.82$ &  & \tabularnewline
${\cal B}_{bc}^{1/2(0)}\to{\cal B}_{bb}^{1/2(1)}l\nu$  & $1.4\times10^{-14}$  & $1.08$ & $1.9\times10^{-14}$ & 0.95\tabularnewline
${\cal B}_{bc}^{1/2(0)}\to{\cal B}_{bb}^{1/2(1)}\tau\nu$  & $4.0\times10^{-15}$ & $0.82$ &  & \tabularnewline
${\cal B}_{bb}^{1/2(1)}\to{\cal B}_{bc}^{3/2(1)}l\nu$  & $1.2\times10^{-14}$ & $0.89$ & $6.4\times10^{-15}$ & $1.43$\tabularnewline
${\cal B}_{bc}^{1/2(1)}\to{\cal B}_{cc}^{3/2(1)}l\nu$  & $9.1\times10^{-15}$ & $0.92$ & $9.0\times10^{-15}$ & $1.18$\tabularnewline
\hline 
\hline 
\end{tabular}
\end{table}

\section{Conclusions}
\label{sec:conclusions}

In summary, we have constructed a heavy diquark effective theory (HDiET), which satisfies  the 
gloabal heavy quark flavor SU(2) symmetry and electromagnetic U(1) symmetry. Imposing these symmetries,
we constructed the coupling terms where the diquark fields interact with the external weak and
electromagnetic sources. Such coupling terms enable us to obtain the effective diquark transition currents
in the small recoil region. On the other hand, for large recoil, the diquark transition currents are
derived from the matching between QCD and DiET at tree level. Furthermore, we simpilfied DiET as
HDiET in the heavy diquark limit, from which we reduced the form factors of the doubly heavy baryon
transition to only one function $\eta(w)$. The reduced vector matrix element is the same as those derived
by heavy quark-diquark symmetry in earlier works. In addition, we pointed out that
$\eta(w)$ is related with the universal soft function which is proportional to the Isgur-Wise
function of heavy meson decays. Thus we obtained the $q^2$-dependence of $\eta(q^2)$ by assuming
a monopole structure. Finally, the obtained form factors are used to predict the semi-leptonic
decay widths of doubly heavy baryons, and the results are consistent with those derived by
LFQM in the earlier works.

\section*{Acknowledgements}
The authors are very grateful to Dr. Chien-Yeah Seng  and Pei-Lun He for useful discussions. This work is
supported in part by    Natural Science Foundation of China under grant No.  11735010,
11911530088, by Natural Science Foundation of Shanghai under grant No. 15DZ2272100,
the DFG and the NSFC through funds provided to the Sino-German CRC 110 ``Symmetries and the Emergence of
Structure in QCD''.
The work of UGM was also supported by the Chinese Academy of Sciences (CAS) President's International
Fellowship Initiative (PIFI)
(Grant No. 2015VMA076) and by the VolkswagenStiftung (Grant No. 93562).

\end{document}